\documentstyle[12pt]{article}

\newcommand{\xxxcal}[1]{{\cal#1}}

\newcommand{\sect}[1]{\setcounter{equation}{0}\section{#1}}

\def\rf#1{(\ref{eq:#1})}
\def\lab#1{\label{eq:#1}}
\def\nonu{\nonumber}
\def\br{\begin{eqnarray}}
\def\er{\end{eqnarray}}
\def\be{\begin{equation}}
\def\ee{\end{equation}}

\def\lb{\lbrack}
\def\rb{\rbrack}

\def\({\left(}
\def\){\right)}

\newcommand{\ct}[1]{\cite{#1}}
\newcommand{\bi}[1]{\bibitem{#1}}

\relax

\def\Tr{\mathop{\rm Tr}}

\newcommand{\sbr}[2]{\left\lbrack\,{#1}\, ,\,{#2}\,\right\rbrack}

\def\a{\alpha}

\def\b{\beta}

\def\d{\delta}
\def\D{\Delta}

\def\g{\gamma}
\def\G{\Gamma}

\def\h{{1\over 2}}
\def\l{\lambda}

\def\o{\over}

\def\O{\Omega}

\def\pa{\partial}
\def\pr{\prime}

\def\ra{\rightarrow}
\def\s{\sigma}

\def\tp0{\Theta_{+}^{(0)}}
\def\tm0{\Theta_{-}^{(0)}}

\def\u2{\mid u\mid^2}
\def\ud{u^{\dagger}}
\def\udu{\( 1 + u^{\dagger}\cdot u\)}

\def\vp{\varphi}

\def\vat{\vartheta}

\def\ca{{\cal A}}

\def\cg{{\cal G}}

\def\ck{{\cal K}}

\def\cn{{\cal N}}
\def\cs{{\cal S}}

\def\ctt{{\cal T}}

\def\f#1#2#3 {f^{#1#2}_{#3}}

\def\win1{{\sf w_{1+\infty}}}

\def\Win1{{\sf W_{1+\infty}}}

\def\rlx{\relax\leavevmode}
\def\inbar{\vrule height1.5ex width.4pt depth0pt}
\def\IZ{\rlx\hbox{\sf Z\kern-.4em Z}}
\def\IR{\rlx\hbox{\rm I\kern-.18em R}}
\def\IC{\rlx\hbox{\,$\inbar\kern-.3em{\rm C}$}}
\def\IN{\rlx\hbox{\rm I\kern-.18em N}}
\def\IO{\rlx\hbox{\,$\inbar\kern-.3em{\rm O}$}}
\def\IP{\rlx\hbox{\rm I\kern-.18em P}}
\def\IQ{\rlx\hbox{\,$\inbar\kern-.3em{\rm Q}$}}
\def\IF{\rlx\hbox{\rm I\kern-.18em F}}
\def\IG{\rlx\hbox{\,$\inbar\kern-.3em{\rm G}$}}
\def\IH{\rlx\hbox{\rm I\kern-.18em H}}
\def\II{\rlx\hbox{\rm I\kern-.18em I}}
\def\IK{\rlx\hbox{\rm I\kern-.18em K}}
\def\IL{\rlx\hbox{\rm I\kern-.18em L}}
\def\one{\hbox{{1}\kern-.25em\hbox{l}}}
\def\0#1{\relax\ifmmode\mathaccent''7017{#1}%
B        \else\accent23#1\relax\fi}

\def\NPB#1#2#3{{\sl Nucl. Phys.} {\bf B#1} (#2) #3}

\begin{document}
\begin{titlepage}
\vspace*{-1cm}
\noindent
October 1998 \hfill{IFT-P/067/98} \\
\phantom{bla}
\hfill{\tt hep-th/9810067}

\vspace{.2in}
\begin{center}
{\large\bf Integrable theories in any dimension \\ 
and homogenous spaces}
\end{center}

\vspace{.2in}

\begin{center}
Luiz A. Ferreira 
 and
Erica E. Leite 

\vspace{.5 cm}
\small

\par \vskip .1in \noindent
Instituto de F\'\i sica Te\'orica - IFT/UNESP\\
Rua Pamplona 145\\
01405-900  S\~ao Paulo-SP, BRAZIL

\normalsize
\end{center}

\vspace{.7in}

\begin{abstract}
We construct local zero curvature representations for non-linear sigma models 
on homogeneous spaces, defined on a space-time of any dimension, following a 
recently proposed approach to integrable theories in dimensions higher than
two. We present some sufficient conditions for the existence of integrable
submodels possessing an infinite number of local conservation laws. 
Examples involving symmetric spaces and group manifolds are given. The $CP^N$
models are discussed in detail.

\end{abstract}
\end{titlepage}

\sect{Introduction}
\label{sec:intro}

The development of techniques to study  non-perturbative aspects of physical 
theories is of crucial importance in practically all areas of
Physics. Many open problems in high energy physics can not be studied with
conventional perturbative methods, and they are in fact  related to the 
non-linear character of  the Lorentz invariant field theories describing the
fundamental interactions of Nature.   

It is perhaps correct to say that many of the developments obtained so far in
such  area involve, in one way or the other, soliton solutions. The most
recent and striking examples are the exact results obtained about the  strong
coupling regime of supersymmetric gauge theories \ct{duality}. They involve a
new version of the electromagnetic duality \ct{olive} which interchanges the
role played by  the two types of fundamental particles of the theory, namely
the excitations of the weakly coupled fields (gauge and matter particles) and 
the solitons (magnetic monopoles and dyons). 

One of the main features of such duality is that the solitons involved saturate
a lower bound for the mass, the so-called Bogomolny bound \ct{bogo}. The
classical solutions for these monopoles can be calculated exactly because they
satisfy some self-duality first order differential equations known as the
Bogomolny-Prasad-Sommerfield (BPS) equations \ct{bogo}. They define a kind of
{\em integrable 
submodel} of the full theory, which present very interesting properties. They
are the couterpart in Minkowski space-time of the  self-duality
condition for the Euclidean Yang-Mills theory containing the instanton
solutions.    

In order to develop techniques to study those types of phenomena one needs a
deep  understanding of the structures and symmetries of the corresponding
theories. However, 
it is well known that soliton solutions are associated to integrability
properties of the model, like infinite number of conservation laws and exact
integration of the equations of motion. In two dimensional space-time, such
relationship is now quite well understood and several techniques have been
developed, based specially on the zero curvature or Lax pair equation for the
theory. Therefore, it is of great importance to attempt to understand the
non-perturbative aspects of non-linear field theories relevant for high energy
physics, like gauge theories, using their integrability properties. 

Recently, it has been proposed a new approach to construct and study 
integrable theories on a space-time of any dimension \ct{afg}. The central 
point of that approach  
is to generalize the zero curvature condition in two dimensions guided by the 
fact that it embodies conservation laws. The extension of integrability
concepts to higher dimensions is a long standing problem. The main 
difficulties are associated to non locality issues that rise when dealing with
higher rank connections. Those problems can
be circunvented by the introduction of auxiliary connections that allow for 
parallel transport. Indeed, it has been shown in \ct{afg} how to obtain {\em
local} zero curvature conditions in space-time of any dimension. The self-dual
Yang-Mills theory and the BPS sector of spontaneously broken gauge theories,
discussed above, have been shown to be examples of theories admiting such
local zero curvature representations. 

One of the interesting aspects of \ct{afg} is that many theories 
presenting the local  zero curvature are not integrable in the sense of
possessing an  
infinite number of conservation laws. However, some of those theories 
contain {\em integrable submodels} that do present an infinite number of
conserved currents. 

The aim of the present paper is to clarify some sufficient conditions for the 
appearance of such integrable submodels. For that, we study Lorentz invariant
field 
theories in space-time of any dimension, defined on homogeneous
spaces. Basically, we treat the non-linear sigma models on coset spaces $G/K$,
and show how to construct the local zero curvature representation for them
using the approach of \ct{afg}. We argue that the equations of motion are
determined by the representation $R^{\cs}$ of the subgroup $K$ defined by the
tangent 
space of $G/K$. The construction of integrable submodels is then shown to be
related to the representations of $G$ which contain $R^{\cs}$ in their
branching in terms of representations of $K$. The submodel is in fact
determined by the constraints that the zero curvature condition, based on those
representations, imposes on the original theory. The number of conservation
laws 
of the submodel is in fact equal to the sum of the dimensions of the
representations of $G$ containing $R^{\cs}$ and leading to the same set of
constraints. In many cases, the number of conserved currents is infinite. 

The paper is organized as follow. In section \ref{sec:summary} we summarize
the ideas involved in the approach of \ct{afg} to integrable theories in any
dimension. In section \ref{sec:coset} we construct the zero curvature
representation for the models defined on coset spaces $G/K$. The conditions for
the 
existence of integrable submodels are discussed in section
\ref{sub:submodels}. The singlets of the subgroup $K$ play an important role
in the construction of such submodels and their conservation laws. That is 
discussed subsection \ref{sub:singlet}. The coset spaces which are symmetric
spaces are considered in section \ref{sec:symsp}. The cases of the group
manifold 
and non-compact symmetric spaces are studied in sections \ref{sec:group} and
\ref{sec:noncompactss} respectively, with some explicit examples
given. Finally, the $CP^N$ models are presented in great detail in section
\ref{sec:cpn}, with the construction of their submodels and corresponding
conservation laws. 

We point out that the criteria for the construction of integrable submodels
discussed here does not exhaust all possibilities. However, we believe it
points towards some very relevant and interesting structures that are
certainly important for the study of integrable theories in higher
dimensions. In particular, the constraints leading to the submodels can
perhaps have an interpretation as a self-duality condition for the full
theory.

\sect{The approach to integrable theories in any \\ di\-men\-sion}
\label{sec:summary}

The central point of the approach of \ct{afg}  
is to generalize the zero curvature condition in two dimensions guided by the 
fact that it embodies conservation laws. Indeed, consider a connection 
$A_{\mu}$ and a curve $\Gamma$ on a two dimensional space time, and define 
the quantity $W$ through the equation
\begin{equation}
{d W\o{d\s}} + A_{\mu} {d x^{\mu}\o{d\s}} W = 0
\lab{weqintro}
\end{equation} 
with $\s$ parametrizing $\G$. Then the zero curvature condition 
\begin{equation}
[\partial_0 + A_0 , \partial_1 + A_1 ]=0 \;.
\lab{2dint}
\end{equation}
is the sufficient condition for the quantity $W$ to be path independent as 
long as its end points are kept fixed. Therefore, if suitable boundary 
conditions are imposed on the fields, like periodic ones where space-time 
can be taken as ${\bf R} \times S^1$ for 
instance, then any power $N$ of the path ordered exponential $\mbox{Tr} 
\left(P\exp (\int_{S^1} A_x (x,t)dx)\right)^N$ is conserved in time.  

The basic idea in \ct{afg} to bring such concepts to higher dimensions, is 
to introduce quantities integrated over hypersurfaces and to find the 
conditions for them to be independent of deformations of the hypersurfaces 
which keep their boundaries fixed. Such an approach will certainly lead to 
conservation laws in a manner very similar to the two dimensional case. 
However, the main problem of that it is how to introduce non-linear zero 
curvatures keeping things as local as possible. The way out is to introduce 
auxiliary connections to allow for parallel transport. The number of 
possibilities  of implementing those ideas increase with the dimensionality 
of space-time. However, the simplest scenario is that where, in a space-time 
of dimension $d+1$, one introduces a rank $d$ antisymmetric tensor 
$B_{\mu_1\mu_2 \ldots \mu_d}$ and a vector $A_{\mu}$.   
The idea can perhaps be best stated using a formulation in ``loop space''. 
On a $d+1$ dimensional space-time $M$ one considers the space 
$\Omega^{d-1}(M,x_0)$ of $d-1$ dimensional closed hypersurfaces based at a 
fixed point $x_0 \in M$. One then introduces on such ``higher loop space'' 
a  $1$-form $\xxxcal{A}$ which is basically the quantity  
$W^{-1}B_{\mu_1\mu_2 \ldots \mu_d}W$ integrated over the closed hypersurfaces 
(see \ct{afg} for details). The quantity $W$ is defined in terms of the 
vector $A_{\mu}$ through \rf{weqintro}. However, for $W$ to be independent of 
the way one integrates it from $x_0$ to a given point on the hypersurface, 
one has to assume that $A_{\mu}$ is flat, i.e.
\be
F_{\mu\nu} = [\partial_{\mu} + A_{\mu} , \partial_{\nu} + A_{\nu} ]=0 \; ; 
\qquad \mu , \nu = 0,1,2 \ldots d 
\lab{stzc}
\ee

Roughly speaking a $d$ dimensional 
closed hypersurface in $M$, based at $x_0$, corresponds to a (one 
dimensional) loop in $\Omega^{d-1}(M,x_0)$. Therefore, the condition to have 
things independent of deformation of hypersurfaces translates in such 
``higher loop space'' to the zero curvature condition for $\xxxcal{A}$, namely 
\be
\xxxcal{F} = 
\delta\xxxcal{A} + \xxxcal{A}\wedge\xxxcal{A} = 0
\lab{loopzc}
\ee
The relation \rf{loopzc} (together with \rf{stzc}) is the generalization of 
the zero curvature \rf{2dint} to higher 
dimensions proposed in \ct{afg}. Although \rf{loopzc} is local in 
$\Omega^{d-1}(M,x_0)$, it is highly non-local in the space-time $M$. Again in 
\ct{afg} it is presented some basic manners of introducing {\em local} 
conditions which are sufficient for the vanishing of $\xxxcal{F}$. The 
relevant local conditions for the applications in this paper are the 
following. 

Let $\cg$ be a Lie algebra and $R$ be a representation of it.  
We introduce the nonsemisimple Lie algebra $\cg_R$ as
\br
\lb T_a \, , \, T_b \rb &=& f_{ab}^c T_c \nonu\\
\lb T_a \, , \, P_i \rb &=& P_j R_{ji}\( T_a\) \nonu\\
\lb P_i \, , \, P_j \rb &=& 0
\lab{rt}
\er
where $T_a$ constitute a basis of $\cg$ and $P_i$ a basis for the abelian
ideal $P$ (representation space). The fact that $R$ is a matrix
representation, i.e.
\begin{equation}
\lb R\( T_a \) \, , \, R\( T_b\)  \rb =  R\( \lb T_a \, , \, T_b \rb \)
\lab{rep}
\end{equation}
follows from the Jacobi identities.

We take the connection $A_{\mu}$ to be in $\cg$ and the rank $d$ 
antisymmetric tensor $B_{\mu_1\mu_2 \ldots \mu_d}$ to be in $P$, i.e.
\begin{equation}
A_{\mu} = A_{\mu}^a T_a \; , \qquad B_{\mu_1\mu_2 \ldots \mu_d} = 
B_{\mu_1\mu_2 \ldots \mu_d}^i P_i
\end{equation}

Then a set of sufficient {\em local} conditions for the vanishing of the 
curvature $\xxxcal{F}$ in \rf{loopzc} is given by 
\be
D_{\mu} {\tilde B}^{\mu} = 0 \; ; \qquad F_{\mu\nu} = 0 
\lab{localzc}
\ee
where we have introduced  the covariant derivative
\be
D_{\mu} \cdot \equiv \partial_{\mu}\, \cdot  + [A_{\mu} \, , \, \cdot \, ]
\lab{covder}
\ee
and the dual of $B_{\mu_1\mu_2 \ldots \mu_d}$ as 
\be
{\tilde B}^{\mu} \equiv {1\o d!} \, 
\varepsilon^{\mu \mu_1\mu_2 \ldots \mu_{d}} \, B_{\mu_1\mu_2 \ldots \mu_d}
\lab{dual}
\ee

The relations \rf{localzc} constitute the {\em local} generalization to higher 
dimensions of the zero curvature condition \rf{2dint}. They lead to local 
conservation laws. Indeed, since the connection $A_{\mu}$ is flat it can be 
written as 
\be
A_{\mu} = -\partial_{\mu} W \, W^{-1}
\lab{puregauge}
\ee
and consequently \rf{localzc} imply that the currents
\be
J_{\mu} \equiv  W^{-1}\, {\tilde B}^{\mu} \, W 
\lab{currents}
\ee
are conserved
\be
\partial_{\mu} \, J^{\mu} = 0
\lab{conserv}
\ee

The zero curvature conditions \rf{localzc} are invariant under the gauge 
transformations
\br
A_{\mu} &\ra & 
g \, A_{\mu} \, g^{-1} - \pa_{\mu} g \, g^{-1} \nonu\\
{\tilde B}_{\mu} &\ra &  
g \, {\tilde B}_{\mu} \, g^{-1} 
\lab{gauge}
\er
and 
\br
A_{\mu} &\ra & A_{\mu} \nonu\\
{\tilde B}_{\mu} &\ra & {\tilde B}_{\mu} + 
\varepsilon_{\mu\mu_1 \ldots \mu_d} D^{\mu_1} \a^{\mu_2 \ldots \mu_d} \equiv 
{\tilde B}_{\mu} + D^{\nu} {\tilde \a}_{\mu\nu}
\lab{newgauge}
\er
where we have introduced the dual ${\tilde \a}_{\mu\nu} \equiv 
\varepsilon_{\mu\nu\mu_2 \ldots \mu_d}\a^{\mu_2 \ldots \mu_d}$. 
In \rf{gauge} $g$ is an element of the group obtained by exponentiating the 
Lie algebra $\cg$. The transformations \rf{newgauge} are symmetries of 
\rf{localzc} as a consequence of the fact that the connection $A_{\mu}$ is 
flat, i.e. $\lb D_{\mu} \, , \, D_{\nu}\rb = 0$. In addition, the parameters 
$\a^{\mu_1 \ldots \mu_{d-1}}$ take values in the abelian ideal $P$. 

The currents \rf{currents} are invariant under the transformations 
\rf{gauge}, but under \rf{newgauge} they transform as 
\be
J_{\mu} \ra J_{\mu} + 
\varepsilon_{\mu\mu_1 \ldots \mu_d} \partial^{\mu_1}\( W^{-1} \, 
\a^{\mu_2 \ldots \mu_d}\, W \) = J_{\mu} + 
\partial^{\nu}\( W^{-1} \,{\tilde \a}_{\mu\nu} \, W \) 
\lab{curtransf}
\ee

The transformations \rf{gauge} and \rf{newgauge} do not commute and their 
algebra is isomorphic to the non-semisimple algebra $\cg_R$ introduced 
in \rf{rt}.

\sect{Integrable theories on coset spaces}
\label{sec:coset}

Consider a Lie group $G$ with Lie algebra $\cg$ and a subgroup $K$ with Lie 
algebra $\ck$. Then we have the decomposition
\be
\cg = \cs + \ck
\lab{deccoset}
\ee
where we have denote by $\cs$ the orthogonal complement of $\ck$ in $\cg$. 
We then have
\be
\sbr{ \ck}{\ck} \subset \ck \quad \sbr{\ck}{\cs} \subset \cs 
\quad \sbr{\cs}{\cs } \subset \cs + \ck
\lab{cosetcom}
\ee

We shall denote by $\Pi$ and $(1-\Pi )$ the orthogonal projections of $\cg$ 
onto $\cs$ and $\ck$ respectively 
\be
\Pi \; : \; \cg \ra \cs \qquad (1 -\Pi ) \; : \; \cg \ra \ck
\lab{cosetproj}
\ee

We are interested in defining models on the coset space $G/K$. The fields of 
such models will be taken to be a set of local coordinates $\zeta^i$ on 
$G/K$, $i=1,2, \ldots {\rm dim} G/K$. 
Locally one can think of $G$ as the direct product of $G/K$ and $K$ and 
therefore a set of local coordinates on $G$ can be taken as the coordinates 
$\zeta^i$ of $G/K$ and some set of local coordinates on $K$. 

We shall consider theories on a $d+1$ dimensional space-time $M$, with 
coordinates $x^{\mu}$, $\mu = 0,1, \ldots d$, and 
therefore  the fields $\zeta^i$ will be mappings from $M$ to $G/K$. 

Following \rf{rt} let us introduce a non-semisimple Lie algebra constructed 
out of $\cg$ and its adjoint representation
\br
\lb T_a \, , \, T_b \rb &=& f_{ab}^c T_c \nonu\\
\lb T_a \, , \, P^{\psi}\(T_b\) \rb &=&  f_{ab}^c \, P^{\psi}\( T_c \) \nonu\\
\lb P^{\psi}\( T_a\) \, , \, P^{\psi}\( T_b\) \rb &=& 0
\lab{rtcoset}
\er
with $T_a$ being a basis for $\cg$ and $P^{\psi}$ denotes the vector space of 
the adjoint representation (where the highest weight is the highest root 
$\psi$ of $\cg$, $R^{\psi}_{cb}\( T_a\) = f_{ab}^c$). 

Let us denote by $S_i$ and $K_r$ the generators of the subspace $\cs$ and 
subalgebra $\ck$ respectively ($i=1,2,\ldots {\rm dim} G/K$, $r=1,2,\ldots 
{\rm dim} K$). We then introduce the potentials
\br
A_{\mu} &\equiv& g^{-1} \pa_{\mu} g =  
g^{-1} {\pa g \o \pa \zeta^i} \, {\pa \zeta^i \o \pa x^{\mu}} 
\equiv A_{\mu}^a T_a \nonu\\
{\tilde B}_{\mu} &\equiv&   P^{\psi}\( \Pi \( g^{-1} \pa_{\mu} g \)\) 
= A_{\mu}^i \,  P^{\psi}\( S_i \)
\lab{cosetpot}
\er
where $g$ is an element of $G$. 

Since the connection $A_{\mu}$ is ``pure gauge'', the flatness condition 
$F_{\mu\nu}=0$ in \rf{localzc} is automatically satisfied. Therefore,  in 
order to get the  local zero curvature conditions, 
we have just to impose that the covariant divergence of ${\tilde B}_{\mu}$ 
vanishes. That will be taken as the equations of motion of our field theory 
on $G/K$. Indeed, the number of such equations of motion is equal to the 
number of fields $\zeta^i$, i.e the dimension of $G/K$. 
So, one gets
\br
D^{\mu} {\tilde B}_{\mu} =  
P^{\psi}\( \Pi \( \pa^{\mu} \( g^{-1} \pa_{\mu} g \)\)\) + 
\sbr{ \( 1-\Pi\)\(g^{-1} \pa^{\mu} g\)}
{P^{\psi}\( \Pi \( g^{-1} \pa_{\mu} g \)\)} =0 
\lab{divbmu}
\er
where, since we are working with the adjoint representation, we have used the 
fact that 
\be
\sbr{\Pi\( g^{-1} \pa^{\mu} g\)}
{P^{\psi}\( \Pi \( g^{-1} \pa_{\mu} g \)\)} =  A^{\mu , i}\, A_{\mu}^j 
\sbr{S_i}{P^{\psi}\( S_j\)} =  A^{\mu , i}\, A_{\mu}^j 
P^{\psi}\(\sbr{S_i}{ S_j} \) = 0
\lab{adjid}
\ee

The action corresponding to \rf{divbmu} is 
\be
S = \h \, \int d^{d+1}x \, \Tr\( \Pi\( g^{-1} \pa_{\mu} g \)\)^2 
= \h \, \int d^{d+1}x \, A_{\mu}^i \, A^{j,\mu} \, \Tr \( S_i S_j \)
\lab{cosetaction}
\ee

Eq. \rf{divbmu} can be written as
\be
\(\pa^{\mu} A_{\mu}^i + A^{\mu , r} A_{\mu}^j R^{\cs}_{ij}\( K_r\) \) 
P^{\psi}\( S_i\) = 0
\lab{coseteqmov}
\ee
where $R^{\cs}_{ij}\( K_r\)$ are the matrices of the representation of the 
subalgebra $\ck$ defined by the subspace $\cs$
\be
\sbr{K_r}{P^{\psi}\( S_j\)}= P^{\psi}\( S_i\) R^{\cs}_{ij}\( K_r\)
\lab{relrep}
\ee

In fact,  the adjoint representation $R^{\psi}$ of $\cg$ decomposes, in
terms of representations of the subalgebra $\ck$, as 
\be
R^{\psi} = R^{\cs} + R^{\ck}
\lab{brancadj}
\ee
where $R^{\cs}$ and $R^{\ck}$ are the representations of $\ck$ defined by the
subspaces $\cs$ and $\ck$ respectively. In fact, $R^{\ck}$ is the adjoint of
$\ck$. Notice those are not  necessarily irreducible.

According to \rf{currents}, the conserved currents for such theory are given 
by (comparing \rf{puregauge} and \rf{cosetpot} one sees that 
$W \equiv g^{-1}$)
\be
J_{\mu} = A_{\mu}^i \,  g \, P^{\psi}\( S_i \) \, g^{-1} = 
A_{\mu}^i \,  R^{\psi}_{ai}\( g \) \,  P^{\psi}\( T_a \) \equiv 
J_{\mu}^a  P^{\psi}\( T_a \)
\lab{cosetcur}
\ee

\newpage 

\sect{The construction of integrable submodels}
\label{sub:submodels}

Although the theory defined above possesses a representation in terms of the 
local zero curvature \rf{localzc}, it does not present an infinite number of 
conserved currents. In fact, as shown in \rf{cosetcur} the number of currents 
is equal to the dimension of $G$. Notice however, that the equations of 
motion \rf{coseteqmov} are determined by the branching of the adjoint 
representation of $G$  into representations of the subgroup $K$. More 
precisely, as shown in \rf{relrep}, what counts is the representation of $K$ 
defined by the subspace $\cs$. Therefore,  any representation of $G$ which 
contains, in its branching rule, that representation of $K$ given by $\cs$,  
can be used to write a zero curvature representation for the model. The way to 
implement that is the following. 

Let $R^{\l}$ be a representation\footnote{It does not have to be irreducible} 
of $G$ that when 
decomposed into representations of the subgroup $K$ presents the 
representation $R^{\cs}$ of $K$ defined by the subspace $\cs$ at least 
once, i.e.
\be
R^{\l} = R^{\cs} + {\rm anything}
\lab{rsdecomp}
\ee

Introduce the non-semisimple Lie algebra
\br
\lb T_a \, , \, T_b \rb &=& f_{ab}^c T_c \nonu\\
\lb T_a \, , \, P^{\l}_{\a} \rb &=& P^{\l}_{\b} R^{\l}_{\b \a}\( T_a \) \nonu\\
\lb P^{\l}_{\a} \, , \, P^{\l}_{\b} \rb &=& 0
\lab{rtcosetl}
\er
with $P^{\l}_{\a}$, $\a =1,2,\ldots {\rm dim} R^{\l}$, being a basis of the 
representation space of $R^{\l}$.  

Following \rf{cosetpot}, define the potentials
\br
A_{\mu} &\equiv& g^{-1} \pa_{\mu} g   
\equiv A_{\mu}^a T_a \nonu\\
{\tilde B}_{\mu}^{\l} &\equiv&   A_{\mu}^i \,  P^{\l}_i 
\lab{cosetpotl}
\er
where $P^{\l}_i$ correspond to a basis of the subspace of $R^{\l}$ 
which carries the representation  $R^{\cs}$ of $\ck$ defined by \rf{relrep},
and which transforms exactly as $P^{\psi}\( S_i\)$, i.e. 
\be
\sbr{K_r}{P^{\l}_j}= P^{\l}_i R^{\cs}_{ij}\( K_r\)
\lab{relrep2}
\ee
Notice that if $R^{\cs}$ is reducible one can rescale the basis of each 
irreducible component independently without changing the relation between 
\rf{relrep} and \rf{relrep2}. 

Therefore, one gets 
\be
D^{\mu} {\tilde B}_{\mu}^{\l} = 
\( \pa^{\mu} A_{\mu}^i \,  P^{\l}_i + 
A_{\mu}^r  A_{\mu}^i\sbr{K_r}{P^{\l}_i}\)  + 
A_{\mu}^i A_{\mu}^j\sbr{S_i}{P^{\l}_j}
\lab{divbl}
\ee
Notice that the first two terms on the r.h.s. of \rf{divbl} are identical to 
\rf{divbmu} (or \rf{coseteqmov}) and 
therefore to the equations of motion of the theory on $G/K$ defined above. 
However, contrary to \rf{adjid} which is an identity, the last term on the 
r.h.s. of \rf{divbl} does not vanish in general. 

Therefore, the submodel of \rf{coseteqmov}  defined by the equations 
\br
\pa^{\mu} A_{\mu}^i + A^{\mu , r} A_{\mu}^j R^{\cs}_{ij}\( K_r\) &=& 0 
\lab{sub1} \\
A_{\mu}^i A^{j,\mu}\, \( 
\sbr{S_i}{P^{\l}_j} + \sbr{S_j}{P^{\l}_i}\) &=& 0 
\; \; ; \; \qquad i,j = 1,2, \ldots \mbox{\rm dim $G/K$}  
\lab{sub2}
\er
admits a representation in terms of the zero curvature
\be
D^{\mu} {\tilde B}_{\mu}^{\l} = 0 \qquad F_{\mu\nu} =0
\lab{zclambda}
\ee 
and therefore possesses the conserved currents
\be
J^{\l}_{\mu} \equiv  A_{\mu}^i \,  g \, P^{\l}_i \, g^{-1} = 
P^{\l}_{\a}  \,  R^{\l}_{\a i}\( g \) \, A_{\mu}^i \equiv 
J_{\mu}^{\l ,\a}  P^{\l}_{\a}
\lab{curl}
\ee
where $P^{\l}_{\a}$, $\a = 1,2,\ldots \mbox{\rm dim $R^{\l}$}$, is a basis of
$R^{\l}$. 

Since $S_i$ and $P^{\l}_i$ transform under the same representation $R^{\cs}$
of $\ck$, it follows that 
$\( \sbr{S_i}{P^{\l}_j} + \sbr{S_j}{P^{\l}_i}\)$ transforms under 
$\( R^{\cs}\otimes R^{\cs}\)_{\rm s}$, where the subscript ${\rm s}$ stands
for the symmetric part of the tensor product. Consider now the branchings 
\be
\( R^{\cs}\otimes R^{\cs}\)_{\rm s} = \sum_{\gamma} \, R^{\gamma}\( \ck\)
\lab{branctensor}
\ee
and
\be
R^{\l} = R^{\cs}\( \ck\) + \sum_{\b} \, R^{\b}\( \ck\)
\lab{brancl}
\ee
where $R^{\gamma}\( \ck\)$ and $R^{\b}\( \ck\)$ are irreducible 
representations of $\ck$. 

Since $\( \sbr{S_i}{P^{\l}_j} + \sbr{S_j}{P^{\l}_i}\)$ corresponds to a given 
state in $R^{\l}$, it follows that it will have to vanish whenever such state
belongs to a representation $R^{\gamma}\( \ck\)$ in \rf{branctensor} which
do not appear in \rf{brancl}. 
Consequently,  the constraints \rf{sub2}
on the fields which are really effective are those corresponding to the
representations $R^{\gamma}\( \ck\)$ in \rf{branctensor} which coincide 
with one of the $R^{\b}\( \ck\)$'s in \rf{brancl}.

Consequently, if the group $G$ possesses a number of representations 
$R^{\l}$'s (which may be infinite) fulfiling the following two requirements
\begin{enumerate}
\item The branching of such representations of $G$ into 
representations of $K$ presents, at least once, the representation $R^{\cs}$ 
of $K$ defined by the subspace $\cs$ (see \rf{relrep})
\item The relation \rf{sub2}, in any 
of such representations, implies the same constraints on the fields. In other
words, the representations  $R^{\gamma}\( \ck\)$'s in \rf{branctensor}, 
appearing in the branching of $R^{\l}$ in \rf{brancl}, are the same for all 
$R^{\l}$'s. 
\end{enumerate}
then the submodel defined in \rf{sub1}-\rf{sub2} possesses a number 
of local conserved currents, given by \rf{curl}, equal to the sum of the
dimensions of such representations $R^{\l}$'s.

\subsection{The role of singlet states}
\label{sub:singlet}

We now discuss a very special case where one can easily construct integrable
sub\-mo\-dels with an infinite number of local conservation laws. Suppose that
$G$ 
possesses a representation $R^{\l}$ which when decomposed into representations
of $\ck$ presents $R^{\cs}$ like in \rf{rsdecomp}, but it also presents a
singlet state 
$P^{\l}_{\Lambda}$ of the subalgebra $\ck$, i.e.\footnote{Clearly, for the 
cases where $\ck$ is abelian, $R^{\l}$  decomposes  into singlet states
only. We then require $P^{\l}_{\Lambda}$ to be a charge zero singlet.}  
\be
\sbr{\ck}{P^{\l}_{\Lambda}} = 0
\lab{singletcond}
\ee
By considering  representations which are  tensor products of $R^{\l}$ 
with itself one then obtains several representations of $\ck$ equivalent 
to $R^{\cs}$, which are given by the tensor product of $R^{\cs}$ with copies of
the singlet  $P^{\l}_{\Lambda}$. For instance, in the case of $R^{\l}\otimes
R^{\l}$ one has that $R^{\cs}\otimes P^{\l}_{\Lambda}$ and  
$P^{\l}_{\Lambda}\otimes R^{\cs}$ are equivalent to $R^{\cs}$. Indeed from
\rf{relrep2}  
\be
\sbr{1\otimes K_r + K_r\otimes 1}{P^{\l}_{\Lambda}\otimes P^{\l}_j}= 
\( P^{\l}_{\Lambda}\otimes P^{\l}_i\) \; R^{\cs}_{ij}\( K_r\)
\lab{tensorrs}
\ee
For the case of $\(\otimes R^{\l}\)^n$ any representation of the
form  
$\(\otimes P^{\l}_{\Lambda}\)^l \otimes R^{\cs}\(\otimes
P^{\l}_{\Lambda}\)^{n-l-1}$  is equivalent to
$R^{\cs}$. Therefore, following \rf{cosetpotl} one introduces the potentials 
\br
A_{\mu}^{(n)} &\equiv&  A_{\mu}^a \sum_{l=0}^{n-1} 
\(\otimes 1\)^l \otimes T_a \(\otimes 1\)^{n-l-1}\nonu\\
{\tilde B}_{\mu}^{\l (n)} &\equiv&   A_{\mu}^i \, \sum_{l=0}^{n-1}
c_{n,l}\;  \(\otimes  P^{\l}_{\Lambda}\)^l \otimes P^{\l}_i 
\(\otimes  P^{\l}_{\Lambda}\)^{n-l-1}
\lab{cosetpottensor}
\er
where $c_{n,l}$ are constants. We introduce such constants because, as we have
pointed out below \rf{relrep2}, one can rescale the basis of each irreducible
component of the re\-pre\-sen\-ta\-tions of $\ck$ independently, without
affecting the 
equations of motion. Only the constraints, defining the submodel, are affected
by the constants $c_{n,l}$. 

The  corresponding zero curvature conditions \rf{zclambda} 
leads in this case, to the following equations of motion (see
\rf{sub1}-\rf{sub2}) 
\be
\pa^{\mu} A_{\mu}^i + A^{\mu , r} A_{\mu}^j R^{\cs}_{ij}\( K_r\) = 0 
\lab{sub1tensor}
\ee
and constraints 
\be
A_{\mu}^i A^{j,\mu}\;  
\sbr{\(\sum_{m=0}^{n-1} 
\(\otimes 1\)^m \otimes S_i \(\otimes 1\)^{n-m-1}\)}{\(\sum_{l=0}^{n-1}
c_{n,l}\;  \(\otimes  P^{\l}_{\Lambda}\)^l \otimes P^{\l}_j 
\(\otimes  P^{\l}_{\Lambda}\)^{n-l-1}\)}  = 0  
\lab{sub2tensor}
\ee
with $i,j = 1,2, \ldots \mbox{\rm dim $G/K$}$.  

Therefore, since \rf{sub1tensor} are the same equations as \rf{coseteqmov} we 
have a submodel of the non-linear sigma model on $G/K$. The subclass of
solutions is determined by the constraints \rf{sub2tensor}. 

The conserved currents obtained from the zero curvature are (see \rf{curl})
\br
J^{\l (n)}_{\mu} &\equiv&  A_{\mu}^i \,  \(\otimes g\)^n 
\, \( \sum_{l=0}^{n-1}
c_{n,l}\;  \(\otimes  P^{\l}_{\Lambda}\)^l \otimes P^{\l}_i 
\(\otimes  P^{\l}_{\Lambda}\)^{n-l-1}\)  \(\otimes g^{-1}\)^n  \nonu\\
&=& A_{\mu}^i  \( \sum_{l=0}^{n-1} c_{n,l} 
V_{\a_1}\( g\) \ldots V_{\a_l}\( g\) 
 R^{\l}_{\a_{l+1} i}\( g \)  V_{\a_{l+2}}\( g\) \ldots V_{\a_n}\( g\)\)
 P^{\l}_{\a_1}\otimes \ldots \otimes P^{\l}_{\a_n} \nonu\\
&\equiv&  J_{\mu}^{\l ,(\a_1\ldots \a_n)}\;  
P^{\l}_{\a_1}\otimes  \ldots \otimes P^{\l}_{\a_n}
\lab{curltensor}
\er
where
\be
g \, P^{\l}_{\Lambda}\, g^{-1} = P^{\l}_{\a}\, V_{\a}\( g\)
\lab{vdefsing}
\ee

Consequently, if one can choose the constants $c_{n,l}$ in such a way that
\rf{sub2tensor} imply for any $n$, the same constraints on the model, one has
an infinite number of local conserved currents for the submodel. Notice that
in such case one has
\be
J_{\mu}^{\l ,(\a_1\ldots \a_n)} = \sum_{l=0}^{n-1} c_{n,l}\; 
V_{\a_1}\( g\) \ldots V_{\a_l}\( g\) 
 J_{\mu}^{\l ,\a_{l+1}}\( g \) \, V_{\a_{l+2}}\( g\) \ldots V_{\a_n}\( g\) 
\lab{curltensor2}
\ee
where $ J_{\mu}^{\l ,\a}=A_{\mu}^i R^{\l}_{\a i}$, are the conserved currents
for the case $n=1$. 

Clearly, if there exists additional singlet states satisfying
\rf{singletcond}, one can use them to construct new currents and submodels. In
fact, the relevant algebraic concept here is that of the kernel of the adjoint
action of the subalgebra $\ck$ on the non-semisimple algebra \rf{rtcosetl},
since $P^{\l}_{\Lambda} \in Ker\( Ad_{\ck}\)$. 
We will discuss examples of such construction on the following sections. 

\newpage 

\sect{The case of symmetric spaces}
\label{sec:symsp}

We now consider the coset spaces $G/K$ which are symmetric spaces 
\ct{helgason}. In such 
cases there exists an involutive authomorphism $\s$, $\s^2 =1$, such that 
$K$ is the invariant subgroup. Then, one decomposes the algebra of 
$G$ as in \rf{deccoset} such that $\cs$ correspond to the odd subspace, i.e.  
\be
\s \( \cs \) = - \cs \; \; ; \qquad \s \( \ck \) = \ck
\lab{ssdec}
\ee
Therefore, instead of \rf{cosetcom} one has 
\be
\sbr{ \ck}{\ck} \subset \ck \quad \sbr{\ck}{\cs} \subset \cs 
\quad \sbr{\cs}{\cs } \subset  \ck
\lab{sscom}
\ee

The projection $\Pi$, introduced in \rf{cosetproj}, can now be performed by 
the automorphism $\s$. Indeed, 
$\( 1-\s\)$ and $\( 1+\s\)$ map $\cg$ into $\cs$ and $\ck$ respectively. 

For any element $g \in G$ we define the so called principal variable 
\ct{olsha,forger1} as 
\be
y\( g\)\equiv g \, \s \( g\)^{-1}
\lab{pv}
\ee
One observes that $y\( g k\)= y\( g\)$ for $k\in K$, and so $y\( g\)$ is 
defined on 
the cosets $G/K$. There exists in fact a one to one correspondence between 
the cosets and the variable $y$, and therefore $y$ can be used to parametrize 
$G/K$. Notice that, $\s \( y \) = y^{-1}$. 

The non-linear sigma model on the symmetric space $G/K$, defined on a 
space-time $M$ of dimension $d+1$, is given by the action 
\be
S \equiv \h \int d^{d+1}x \, \Tr \( y^{-1}\pa_{\mu}y\)^2 
\lab{sslag}
\ee
which corresponds to the equations of motion
\be
\pa^{\mu}\( y^{-1}\pa_{\mu}y\) =0
\lab{sseqmov}
\ee

Such theories admit a quite simple representation in terms of the local zero 
curvature conditions \rf{localzc}. Consider the non-semisimple Lie algebra 
\rf{rtcoset} and introduce
\br
A_{\mu} &\equiv& y^{-1} \pa_{\mu} y \nonu\\
{\tilde B}_{\mu}  &\equiv& P^{\psi}\( y^{-1} \pa_{\mu} y\) 
\lab{sspot}
\er
Clearly, $F_{\mu\nu}=0$, and 
the condition $D^{\mu} {\tilde B}_{\mu}=0$ is equivalent to \rf{sseqmov}. 

Notice that
\be
y^{-1} \pa_{\mu} y = \s \( g\) \( g^{-1} \pa_{\mu} g - 
\s \( g^{-1} \pa_{\mu} g \) \) \s \( g\)^{-1}
\ee
Therefore, performing the gauge transformation \rf{gauge} with 
$\s \( g\)^{-1}$ one obtains
\br
A_{\mu} &\ra& g^{-1} \pa_{\mu} g \nonu\\
{\tilde B}_{\mu}  &\ra& P^{\psi}\( \( 1-\s \) \( g^{-1} \pa_{\mu} g\) \) 
\lab{sspot2}
\er
Notice that these potentials are in the same gauge as those in \rf{cosetpot}. 
Therefore, the analysis presented in \rf{divbmu}-\rf{cosetcur}, as well as 
the discussion about integrable submodels in section \ref{sub:submodels},  
hold true in the present case.

\sect{The case of the group manifold}
\label{sec:group}

The non-linear sigma model on a group manifold $G$, defined on a space-time
$M$ of dimension $d+1$, is given by
\be
S \equiv \h \int dx^{d+1} \, \Tr \( g^{-1} \pa_{\mu} g\)^2 \; \; ; 
\qquad \qquad g \in G 
\lab{groupaction}
\ee
and the corresponding equations of motion are
\be
\pa^{\mu} \( g^{-1} \pa_{\mu} g\) = 0 \; \; ; \qquad \mbox{\rm or} \qquad 
\pa^{\mu} \( \pa_{\mu} g g^{-1} \) = 0
\lab{groupeqmov}
\ee

These models have already been studied in \ct{joaq} using the same zero
curvature  
approach proposed in \ct{afg} and some interesting integrable submodels as
well as the corresponding conserved currents were constructed.  
Any group $G$, however, is a symmetric space \ct{forger1} and therefore the
theory \rf{groupaction} can be studied using the techniques of sections 
\ref{sec:coset} and \ref{sec:symsp}. That may help making more systematic the
construction of integrable submodels. 
The relevant symmetric space is $G\otimes
G/G_D$, where the elements of the tensor group $G\otimes G$ are  of
the form $g_1\otimes g_2$, with $g_1 , g_2 \in G$, and $G_D$ is the diagonal
subgroup with elements $g\otimes g$, with $g \in G$. The involutive
automorphism is 
\be
\sigma \( g_1\otimes g_2\) = g_2\otimes g_1
\lab{groupsigma}
\ee
and indeed $G_D$ is the invariant subgroup under $\sigma$. 

The group $G$ is diffeomorphic to  $G\otimes G/G_D$, with the diffeomorphism 
$G\otimes G \ra G$ being given by $g_1\otimes g_2 \ra g_1 g_2^{-1}$. 
Obviously the kernel is $G_D$ itself. 

The principal variable $y$ introduced in \rf{pv} is given by
\be
y \( g_1\otimes g_2 \) = g_1\otimes g_2 \sigma \( g_1\otimes g_2 \)^{-1} = 
g_1 g_2^{-1} \otimes \( g_1 g_2^{-1}\)^{-1}
\ee 

Notice that $y$ is always the tensor product of a given element with its
inverse. Since $y$ parametrizes $G\otimes G/G_D$ and since that has the same
dimension as $G$, one can always choose a gauge where $y = g \otimes g^{-1}$,
with $g \in G$. Therefore, the equation of motion \rf{sseqmov} becomes 
\be
\pa^{\mu}\( y^{-1}\pa_{\mu}y\) = \pa^{\mu}\( g^{-1}\pa_{\mu}g \) \otimes 1 - 
1 \otimes \pa^{\mu}\( \pa_{\mu}g g^{-1} \) = 0 
\lab{groupeqmov2}
\ee
Therefore the non-linear sigma models defined on $G\otimes G/G_D$ and $G$ are
the same, since \rf{groupeqmov} and \rf{groupeqmov2} are equivalent.  

Following \rf{sspot} we can then introduce the potentials $A_{\mu}$ and
${\tilde B}_{\mu}$, which in the gauge \rf{sspot2} are given by 
\br
A_{\mu} &=& p^{-1}\pa_{\mu}p  \otimes 1 - 
1 \otimes \pa_{\mu}p\, p^{-1} = 
\( p^{-1}\otimes p\) 
\pa_{\mu} \( p\otimes p^{-1} \)  \nonu\\
{\tilde B}_{\mu}  &=& 
P^{\psi}\( \( 1-\s \) \( p^{-1}\pa_{\mu}p  \otimes 1 - 
1 \otimes  \pa_{\mu}p\, p^{-1} \) \)
\lab{grouppot}
\er
where $p$ is such that $p\, p=g$, with $g$ being the group
element in the definition of $y$ (indeed $y\(p\otimes p^{-1}\) =
g\otimes g^{-1}$, see \rf{pv} and \rf{groupsigma}). 

The local zero curvature conditions \rf{localzc} then imply the equation of
motion \rf{groupeqmov2}, because $F_{\mu\nu}=0$ is trivially satisfied since
$A_{\mu}$ is of the pure gauge form, and
\be
D^{\mu}{\tilde B}_{\mu} = \( 1-\s \) \( \( 
\pa^{\mu}\( p^{-1}\pa_{\mu}p \) + 
\pa^{\mu}\( \pa_{\mu}p p^{-1} \) + 
\sbr{p^{-1}\pa_{\mu}p}{\pa^{\mu}p p^{-1}}
\)\otimes 1 \) = 0 
\ee
is equivalent to \rf{groupeqmov2}. 

The conserved currents \rf{currents} are given by 
\be
J_{\mu} = \( p\otimes p^{-1}\) {\tilde B}_{\mu} 
\( p^{-1}\otimes p\) = 
\( \pa_{\mu} g g^{-1}\) \otimes 1 - 1 \otimes \( g^{-1} \pa_{\mu} g \)
\ee
which correspond to the Noether currents associted to the invariance of
\rf{groupaction} under the global right and left translations by elements of
$G$. 

Let us now consider the construction of integrable submodels of the theory 
\rf{groupaction}, which possess a larger number of conserved currents, using
the ideas of section \ref{sub:submodels}. 
The algebra $\cg \oplus \cg$ of $G\otimes G$ decomposes under $\s$ as (see
\rf{ssdec}) 
\be
\cg \oplus \cg = \cs + \ck 
\ee
with 
\be
\cs \equiv \{ \ctt^{\cs}_a \equiv 1\otimes T_a - T_a \otimes 1 \} 
\;\; ; \qquad \qquad 
\ck \equiv \{ \ctt^{\ck}_a \equiv 1\otimes T_a + T_a \otimes 1 \}
\ee
where $T_a$, $a=1,2,\ldots {\rm dim}\; G$, are the generators of the algebra
$\cg$ of $G$ ($\sbr{T_a}{T_b} = f_{ab}^c \, T_c$). Therefore
\br
\sbr{\ctt^{\ck}_a}{\ctt^{\ck}_b} = f_{ab}^c \,\ctt^{\ck}_c \; \; ; \qquad 
\sbr{\ctt^{\ck}_a}{\ctt^{\cs}_b} = f_{ab}^c \,\ctt^{\cs}_c \; \; ; \qquad 
\sbr{\ctt^{\cs}_a}{\ctt^{\cs}_b} = f_{ab}^c \,\ctt^{\ck}_c 
\lab{groupsimcom}
\er

In fact, denoting 
\be
p^{-1}\pa_{\mu} p \pm \pa_{\mu} p\, p^{-1}
\equiv A^{\pm}_{\mu} \equiv A^{\pm ,a}_{\mu}\, T_a 
\lab{apmdef}
\ee
one obtains from \rf{grouppot} that
\br
A_{\mu} &=& \h A^{-,a}_{\mu}\, \ctt^{\ck}_a 
 + \h  A^{+,a}_{\mu}\, \ctt^{\cs}_a \nonu \\
{\tilde B}_{\mu} &=&  A^{+,a}_{\mu}\, 
P^{\psi} \( \ctt^{\cs}_a \) 
\lab{grouppot2}
\er

The equations of motion are then written as
\be
\pa^{\mu}  A^{+,a}_{\mu}  + 
\h \, f_{bc}^a  A^{-,b}_{\mu}A^{+,c,\mu} = 0 
\lab{groupeqmot3}
\ee

As we have discussed in section \ref{sub:submodels}, the part of $A_{\mu}$
that really contribute to the equations of motion is that in $\ck$. In
addition, those equations are determined by the representation of $\ck$
defined by the subspace $\cs$. But $\ck$, the algebra of $G_D$,  is
isomorphic to $\cg$ and therefore that representation is the
adjoint. Consequently, as pointed out in \ref{sub:submodels}, any
representation of $G\otimes G$ that contains the adjoint of $G$ in its
branching rule can be used to write a zero curvature for submodels of the
theory 
\rf{groupaction}. Given two representaions $R^{\l}$ and $R^{\l^{\pr}}$ of $G$
one can construct a representation of $G\otimes G$ by taking the tensor
product of them. Therefore, one should look for representations $R^{\l}$ and
$R^{\l^{\pr}}$ such 
\be
R^{\l}\otimes R^{\l^{\pr}} = \mbox{\rm adjoint of $G$} + \mbox{\rm anything} 
\ee
The construction of the zero curvature for submodels of \rf{groupaction} (with
infinite number of conserved currents)  
is done by following the ideas described in  section \ref{sub:submodels}. 

\subsection{The example of $SU(2)$}
\label{sub:groupsu2}

Let us ilustrate those ideas with the example of $SU(2)$  where 
the commutation relations are given by 
\be
\sbr{T_i}{T_j} = i\, \varepsilon_{ijk}\, T_k \;\; ; \qquad \qquad i,j,k=1,2,3 
\lab{su2unitarycom}
\ee

The equations of motion are those of \rf{groupeqmot3} with $f_{ab}^c$ replaced
by $i\, \varepsilon_{ijk}$. 
We now use the fact that the adjoint (triplet) of $SU(2)$ can be obtained by
the tensor product of two doublets, i.e.
\be
{\bf 2} \otimes {\bf 2} = {\bf 3} + {\bf 1}
\ee

Denoting the basis of the doublet by $P^{(1/2)}_{\pm 1/2}$ one has 
($T_{\pm} \equiv T_1 \pm i T_2$) 
\be
\sbr{T_3}{P^{(1/2)}_{\pm 1/2}}= \pm \h \, P^{(1/2)}_{\pm 1/2} \; \; ; \qquad 
\sbr{T_{\pm}}{P^{(1/2)}_{\mp 1/2}}= P^{(1/2)}_{\pm 1/2} 
\ee
For the tensor product representation space we take the basis
\br
P^{(\h , \h )}_1 &\equiv&  i\,\( P^{(1/2)}_{1/2}\otimes P^{(1/2)}_{1/2} - 
P^{(1/2)}_{-1/2}\otimes P^{(1/2)}_{-1/2}\) \nonu\\
P^{(\h , \h )}_2 &\equiv&   P^{(1/2)}_{1/2}\otimes P^{(1/2)}_{1/2} + 
P^{(1/2)}_{-1/2}\otimes P^{(1/2)}_{-1/2} \nonu\\
P^{(\h , \h )}_3 &\equiv&  -i\,\( P^{(1/2)}_{1/2}\otimes P^{(1/2)}_{-1/2} + 
P^{(1/2)}_{-1/2}\otimes P^{(1/2)}_{1/2}\) \nonu\\
P^{(\h , \h )}_{\Lambda} &\equiv&  
 P^{(1/2)}_{-1/2}\otimes P^{(1/2)}_{1/2} -
P^{(1/2)}_{1/2}\otimes P^{(1/2)}_{-1/2} 
\lab{tensorbasissu2}
\er
One can check that they satisfy
\br
\sbr{\ctt^{\ck}_i}{P^{(\h , \h )}_j} &=& 
i\, \varepsilon_{ijk}\,P^{(\h , \h )}_k \; \; \qquad 
\sbr{\ctt^{\cs}_i}{P^{(\h , \h )}_j} =
i\, \d_{ij}\,P^{(\h , \h )}_{\Lambda} \nonu\\
\sbr{\ctt^{\ck}_i}{P^{(\h , \h )}_{\Lambda}} &=& 0 
\; \; \; \qquad \qquad \qquad 
\sbr{\ctt^{\cs}_i}{P^{(\h , \h )}_{\Lambda}} = 
-i\,P^{(\h , \h )}_{i}
\lab{comtphh}
\er

We then introduce the potential
\br
{\tilde B}_{\mu}^{(\h , \h )} &\equiv& A^{+,i}_{\mu} \,P^{(\h , \h )}_i
\lab{bpottensorsu2}
\er
which, like \rf{grouppot2}, contains the states transforming under the adjoint
(triplet).  

One can easily verify that the equation 
\be
D^{\mu} {\tilde B}_{\mu}^{(\h , \h )} = 0
\ee
with the same potential $A_{\mu}$ as in \rf{grouppot2}, gives the same
equations of motion  \rf{groupeqmot3}. \\ 
However, it has a component in the
direction of the singlet state which imposes the following
constraint on the model
\br
A^{+,i}_{\mu}\, A^{+,j,\mu} \sbr{\ctt^{\cs}}{P^{(\h , \h )}_j} = 
i \, A^{+,i}_{\mu}\, A^{+,i,\mu} \, P^{(\h , \h )}_{\Lambda} = 0 \; \; \quad 
\ra \quad A^{+,i}_{\mu}\, A^{+,i,\mu} = 0 
\er

Using \rf{apmdef}, and the fact that 
$g^{-1}\pa_{\mu}g = p^{-1}\( p^{-1}\pa_{\mu}p + \pa_{\mu}p\, p^{-1} \)\, p$,  
such constraint can be written as
\be
\Tr \( p^{-1}\pa_{\mu}p + \pa_{\mu}p\, p^{-1} \)^2 = 
\Tr \( g^{-1}\pa_{\mu}g \)^2 = 0 
\ee
where we have used the fact that $\Tr \( T_i T_j\) \sim \d_{ij}$. 
Therefore, such constraint implies that the action \rf{groupaction}
vanishes when evaluated on the solutions of such submodel.

The corresponding four conserved currents are 
\br
J_{\mu}^{(\h , \h )} = \( p\otimes p^{-1}\) \, 
{\tilde B}_{\mu}^{(\h , \h )}\, \( p^{-1}\otimes p\)  = 
J_{\mu}^{(\h , \h ),\a \b} \, P^{(1/2)}_{\a}\otimes P^{(1/2)}_{\b}
\er
with $\a , \b = \pm 1/2$, and 
\br
J_{\mu}^{(\h , \h ),\a \b} &\equiv& 
\( A_{\mu}^{+,2} + i A_{\mu}^{+,1}\) D_{\h}^{\a}\( p\) D_{\h}^{\b}\( p^{-1}\) 
+ \( A_{\mu}^{+,2} - i A_{\mu}^{+,1}\) 
D_{-\h}^{\a}\( p\) D_{-\h}^{\b}\( p^{-1}\) \nonu\\
&-& i  A_{\mu}^{+,3} \(  D_{\h}^{\a}\( p\) D_{-\h}^{\b}\( p^{-1}\) + 
D_{-\h}^{\a}\( p\) D_{\h}^{\b}\( p^{-1}\)\)
\er
and where
\be
p \,  P^{(1/2)}_{\a} \, p^{-1} =  P^{(1/2)}_{\b}\,  D_{\a}^{\b}\( p\)
\ee
 
Since we have a $\ck$-singlet in such representation we can use the ideas of
section \ref{sub:singlet} to construct submodels with larger conservation
laws.  We then introduce the potentials 
\br
A_{\mu}^{(n)} &\equiv&  \h A_{\mu}^{-,i}\, \sum_{l=0}^{n-1} 
\(\otimes 1\)^l \otimes \ctt^{\ck}_i \(\otimes 1\)^{n-l-1}
+ \h A_{\mu}^{+,i}\, \sum_{l=0}^{n-1} 
\(\otimes 1\)^l \otimes \ctt^{\cs}_i \(\otimes 1\)^{n-l-1}
\nonu\\
{\tilde B}_{\mu}^{(\h , \h ),(n)} &\equiv&   A_{\mu}^{+,i} \, \sum_{l=0}^{n-1}
c_{n,l}\;  \(\otimes  P^{{(\h , \h )}}_{\Lambda}\)^l \otimes 
P^{(\h , \h )}_i 
\(\otimes  P^{(\h , \h )}_{\Lambda}\)^{n-l-1}
\lab{cosetpottensorsu2}
\er

As we have argued in section \ref{sub:singlet} the zero curvature condition
for these potentials give the same equations of motion as those of
\rf{bpottensorsu2}, i.e. \rf{groupeqmot3}. However, the constraints correspond
to those  given in \rf{sub2tensor}. One can easily check that for the case
$n=2$ no constraints are imposed on the fields if we choose
\be
c_{2,0} + c_{2,1} = 0
\ee
However, for $n>2$ there are no choices for $c_{n,l}$ which can make the
constraints weaker. In such cases we have to impose
\be
A^{+,i}_{\mu}\, A^{+,j,\mu} = 0 \; \; \qquad \qquad \mbox{\rm for any $i,j$}
\lab{strongconstr}
\ee

Denoting the parameters of the group by $\zeta^i$, $i=1,2,3$, one gets from
\rf{apmdef} that 
$A^{+,i}_{\mu} = {{\cal M}^{(+)}}^{i}_j \( p\) \pa_{\mu} \zeta^j$,
with ${{\cal M}^{(+)}}^{i}_j \( p\) $ being an invertible matrix. 
Therefore, the
constraints $\rf{strongconstr}$ imply that 
$\pa_{\mu} \zeta^i \pa^{\mu} \zeta^j=0$. Now, one can also write 
$A^{-,i}_{\mu} = {{\cal M}^{(-)}}^{i}_j \( p\) \pa_{\mu} \zeta^j$, and
consequently $A^{-,i}_{\mu}\, A^{+,j,\mu} = 0$. In addition,
\be 
\pa^{\mu}A^{+,i}_{\mu}= {{\cal M}^{(+)}}^{i}_j \( p\) \pa^2  \zeta^j + 
\pa_k \( {{\cal M}^{(+)}}^{i}_j\) \pa^{\mu} \zeta^k \pa_{\mu} \zeta^j
\ee 
Therefore, from the equations of motion \rf{groupeqmot3} and constraints 
\rf{strongconstr}, we get that the submodel is defined by 
\be
\pa^2  \zeta^j =0 \; \; ; \qquad \qquad \pa_{\mu} \zeta^i \pa^{\mu} \zeta^j=0
\lab{cp3su2}
\ee

If one allows the fields to be complex (i.e. work with $SL(2,\IC )$), then
\rf{cp3su2} is the same as that submodel of $CP^3$ we discuss in section
\ref{subsub:singletcpn}. The conserved currents can be evaluated using
\rf{curltensor2}. However, we do not discuss them in more detail here because
we shall treat such type of submodel in section \ref{subsub:singletcpn}.

\sect{The example of non-compact symmetric spaces}
\label{sec:noncompactss}

We now consider the symmetric spaces $G/K$ where $G$ is a real 
non-compact simple 
Lie group furnished with a Cartan involution $\s$, $\s^2 =1$, with $K$,   
invariant under $\s$, being the maximal compact subgroup of $G$ 
\ct{helgason}. The Cartan property of $\s$ means that if $\Tr$ is a 
$\s$-invariant bilinear form for the algebra $\cg$ of $G$ 
($\Tr\( TT^{\pr}\) = 
\Tr\(\s\( T\) \s\( T^{\pr}\)\)$) then $\Tr\( T\s\( T^{\pr}\) \)$ is negative 
definite. That implies that $\Tr\( TT^{\pr}\)$ is: {\it i)} positive definite 
if $T , T^{\pr}\in \cs$; {\it ii)} negative definite if $T , T^{\pr}\in \ck$; 
and {\it iii)} zero if $T\in \cs$ and $T^{\pr}\in \ck$. 

Such symmetric spaces have some very special properties due to the so-called 
Iwasawa decomposition of $G$ \ct{helgason}. Let $\ca$ denote the maximal 
abelian subspace of $\cs$. It then follows that the adjoint action of $\ca$ 
in $\cg$ can be simultaneously diagonalized. We denote $\cg_{\g} \equiv 
\{ T \in \cg \mid \sbr{H}{T}= \g\( H\) T \; , \; 
{\mbox{\rm for $H\in \ca$}}\}$. 
We now define the nilpotent subalgebra $\cn \equiv \sum_{\g > 0} \cg_{\g}$. 
The Iwasawa decomposition corresponds to \ct{helgason}
\be
\cg = \cn + \ca + \ck \; \; ; \qquad g = n \, a \, k \equiv b \, k 
\lab{iwasawa}
\ee
where $k\in K$, and $n$ and $a$ are elements of the subgroups obtained by 
exponentiating $\cn$ and $\ca$ respectively, and where we have introduced 
$b \equiv na$. 

It then follows that such symmetric spaces are endowed with a hidden group 
theoretic structure, since the elements of $G/K$ can be put into a one to one 
correspondence with the elements $b$ of the so-called Borel subgroup. 
Even though $G/K$ is a not a group itself, one can parametrize it 
by the group elements $b$. 

Using the symmetry \rf{gauge} one can choose a gauge where the potentials 
\rf{sspot2}, in the case of such non-compact symmetric spaces, take  the 
form
\br
A_{\mu} &=& b^{-1} \pa_{\mu} b = a^{-1} \pa_{\mu} a + 
a^{-1}\( n^{-1} \pa_{\mu}n \) a \lab{sspot3} \\
{\tilde B}_{\mu}  &=& P^{\psi}\( \( 1-\s \) \( b^{-1} \pa_{\mu} b\) \) = 
 P^{\psi}\( 2 \,  a^{-1} \pa_{\mu} a + a^{-1}\( n^{-1} \pa_{\mu}n \) a - 
a\s\( n^{-1} \pa_{\mu}n \) a^{-1}\)  \nonu
\er
where we have used the fact that $\s\( a\) = a^{-1}$, since $\ca \in \cs$ and 
so $\s\( \ca\) = - \ca$.

\subsection{The case where $G$ is the normal real form}
\label{subsec:normal}

Consider the case where the algebra $\cg$ of $G$ is spanned by real linear 
combinations of the Chevalley basis $H_a$, 
$a=1,2, \ldots \mbox{\rm rank $\cg$}$, $E_{\a}$ and $E_{-\a}$, with $\a$ 
being the positive roots of $\cg$. That is the maximally non-compact real form 
of the corresponding complex simple Lie group, and its called the normal 
form. The Cartan involution we consider is given by 
($\s^2=1$)
\be
\s\( H_a \) = - H_a \; ; \qquad \s\( E_{\a} \) = - E_{-\a}
\lab{carinv}
\ee

Therefore 
\br
\cs &=& \{ H_a , \; a=1,2, \ldots \mbox{\rm rank $\cg$}\; ; \; 
 E_{\a} + E_{-\a}, \; \mbox{\rm for any positive root $\a$}\} \nonu\\
\ck &=& \{  E_{\a} - E_{-\a}, \; \mbox{\rm for any positive root $\a$}\}
\er
and
\be
\ca = \{ H_a , \; a=1,2, \ldots \mbox{\rm rank $\cg$}\} \; ; \qquad 
\cn = \{ E_{\a}, \; \mbox{\rm for any positive root $\a$}\} 
\ee

Parametrizing the group elements as
\be
a= \exp \( -\h\, \sum_{a=1}^{{\rm rank}\, \cg} \vp^a \, H_a \) \; ; \qquad 
n= \exp \( \sum_{\a >0} \zeta^{\a} \, E_{\a} \)
\ee
one gets from \rf{sspot3}
\br
A_{\mu} &=&  -\h\,\sum_{a=1}^{{\rm rank}\, \cg} \pa_{\mu} \vp^a \, H_a + 
\sum_{\a ,\b >0} \pa_{\mu} \zeta^{\a}\, {\cal V}_{\a\b}\( \zeta \)\, 
e^{\h K_{\b a}\vp^a} \, E_{\b}
\lab{sspot4}\\
{\tilde B}_{\mu}  &=& 
P^{\psi}\( -\sum_{a=1}^{{\rm rank}\, \cg} \pa_{\mu} \vp^a \, H_a + 
\sum_{\a ,\b >0} \pa_{\mu} \zeta^{\a}\, {\cal V}_{\a\b}\( \zeta \)\, 
e^{\h K_{\b a}\vp^a} \, 
\( E_{\b} + E_{-\b}\) \)
\nonu 
\er
where $ K_{\b a}\equiv {2 \b \cdot \a_a \o \a_a^2}$, with $\a_a$ being a 
simple root of $\cg$, and 
\be
n^{-1}{\pa n \o \pa \zeta^{\a}} \equiv 
\sum_{\b >0} \, {\cal V}_{\a\b}\( \zeta \)\,  E_{\b} 
\ee

The conserved currents \rf{currents} are given by ($W^{-1} \equiv na$)
\br
J_{\mu} &=& na\, {\tilde B}_{\mu} a^{-1} n^{-1} 
\lab{curnormal}\\
&=& P^{\psi}\( n\, \( -\sum_{a=1}^{{\rm rank}\, \cg} \pa_{\mu} \vp^a \, H_a + 
\sum_{\a ,\b >0} \pa_{\mu} \zeta^{\a}\, {\cal V}_{\a\b}\( \zeta \)\, 
 \(  E_{\b} + e^{ K_{\b a}\vp^a} \, E_{-\b}\) \) n^{-1} \) \nonu 
\er

\subsubsection{The example of $sl(2)$}

In such case there is just one positive root, and therefore we denote 
$a=e^{-\h\vp H}$, $n=e^{\zeta E_{+}}$. The 
commutation relations for $sl(2)$ are 
\be
\sbr{H}{E_{\pm}} = \pm 2\, E_{\pm} \; ; \qquad \sbr{E_{+}}{E_{-}}= H 
\lab{sl2com}
\ee
We have $n^{-1}{\pa n \o \pa \zeta}= E_{+}$, and so 
${\cal V}_{\a\b} \equiv 1$. 
Then, from \rf{sspot4} 
one gets
\be
D^{\mu}{\tilde B}_{\mu} = P^{\psi}\(  \,\( -\pa^2 \vp  +  
e^{2\vp}\, \( \pa_{\mu}\zeta\)^2\) H + 
e^{\vp}\, \( \pa^2 \zeta + 2 \pa_{\mu}\vp \pa^{\mu}\zeta \) 
\( E_{+} + E_{-}\)\)
\ee
Consequently, the local zero curvature conditions \rf{localzc} imply the 
equations of motion
\br
\pa^2 \vp  - e^{2\vp}\, \( \pa_{\mu}\zeta\)^2 &=& 0\\
\pa^2 \zeta + 2 \pa_{\mu}\vp \pa^{\mu}\zeta &=& 0
\lab{eqmovsl2}
\er

The conserved currents \rf{curnormal} are given by
\be
J_{\mu} = J_{\mu}^{+}\, P^{\psi}\( E_{+}\) + 
 J_{\mu}^{0}\,P^{\psi}\( H\) +  J_{\mu}^{-}\, P^{\psi}\( E_{-}\)
\ee
with
\br
J_{\mu}^{+}&=& \( 1 - \zeta^2 \, e^{2\vp}\)\, \pa_{\mu} \zeta 
+ 2 \zeta\, \pa_{\mu} \vp \nonu\\
J_{\mu}^{0}&=& -\pa_{\mu} \vp + e^{2\vp}\, \zeta \, \pa_{\mu} \zeta \nonu\\
J_{\mu}^{-}&=& e^{2\vp}\, \pa_{\mu} \zeta 
\lab{cursl2}
\er

Following the discussion of section \ref{sub:submodels}, we now construct a 
submodel of \rf{eqmovsl2} that possesses an infinite number of local 
conserved currents. In the notation of that section, the subgroup $K$ here is 
$SO(2)$ (or $U(1)$) and it is generated by $\( E_{+} - E_{-}\)$. The subspace 
$\cs$ is generated by $H$ and $\( E_{+} + E_{-}\)$. Since those generators do
not diagonalize the action of the $SO(2)$ subgroup, we shall work with the 
basis\footnote{Notice that formally, the $sl(2)$ generated by $T_3$ and 
$T_{\pm}$ is not
the same as that generated by $H$ and $E_{\pm}$, since they are related by
{\em complex} linear combinations. They are in fact distinct real forms of the
same complex $sl(2,\IC )$.}
\be
T_3\equiv {1\o 2i}\, \( E_{+} - E_{-}\)\;\; ; \qquad 
T_{\pm} \equiv \h \, \( H \pm i \( E_{+} + E_{-}\) \) 
\lab{newbasis}
\ee
which satisfy 
\be
\sbr{T_3}{T_{\pm}} = \pm \, T_{\pm} \; ; \qquad \sbr{T_{+}}{T_{-}}= 2\, T_3 
\lab{newsl2com}
\ee

Therefore,  the potentials \rf{sspot4} become 
\br
A_{\mu} &=& -\h \( \pa_{\mu}\vp  + ie^{\vp} \pa_{\mu} \zeta  \) T_{+} - 
\h \( \pa_{\mu}\vp  - i e^{\vp}\pa_{\mu} \zeta  \) T_{-} + 
i e^{\vp}  \pa_{\mu} \zeta T_3 \equiv A_{\mu}^i \, T_i \nonu \\
{\tilde B}_{\mu} &=& 
- P^{\psi}\( \( \pa_{\mu}\vp  + i e^{\vp}\pa_{\mu} \zeta  \) T_{+} + 
\( \pa_{\mu}\vp  - i e^{\vp}\pa_{\mu} \zeta  \)  T_{-}\) 
\lab{newpotsl2}
\er

Obviously, the adjoint of $SL(2)$ possesses a singlet state of the $SO(2)$
subalgebra, namely $P^{\psi}\( T_3 \)$. Therefore, using the ideas of
section \ref{sub:singlet} we can construct submodels with large number of
conservation laws. So, following \rf{cosetpottensor} we introduce
\br
A_{\mu}^{(n)} &\equiv&  A_{\mu}^i \sum_{l=0}^{n-1} 
\(\otimes 1\)^l \otimes T_i \(\otimes 1\)^{n-l-1}
\lab{pottensornoncompact}\\
{\tilde B}_{\mu}^{\psi (n)} &\equiv&    \, \sum_{l=0}^{n-1}
\;  \(\otimes  P^{\psi}\( T_3\)\)^l 
\otimes \(  A_{\mu}^+ P^{\psi}\( T_+\) + 
  A_{\mu}^- P^{\psi}\( T_-\)\)  
\(\otimes  P^{\psi}\( T_3\)\)^{n-l-1}
\nonu 
\er 
According to the comments below \rf{relrep2} and \rf{cosetpottensor} we could
have rescaled the basis of the irreducible representations of $SO(2)$,
independently. However, such freedom would not produce more submodels with
infinite number of conserved currents (see discussion below
\rf{constrainttensor1} for a similar situation)\footnote{There is an
additional choice which would lead to the constraint 
$\mid\pa_{\mu}\vp  + i e^{\vp}\pa_{\mu} \zeta \mid^2 = 0$, instead of
\rf{constrsl2}. However, the method of section \ref{sub:singlet} would lead to
conserved currents for the case $n=2$ only.} 

As we have argued in section \ref{sub:singlet} the zero curvature for these
potentials lead to the equations of motion \rf{eqmovsl2} and the constraints
corresponding to \rf{sub2tensor}. One can verify that such constraints for any
value of $n$ correspond to $\( A_{\mu}^+\)^2 = \( A_{\mu}^-\)^2 = 0$, which is
equivalent 
to 
\be
\( \pa_{\mu}\vp  + i e^{\vp}\pa_{\mu} \zeta  \)^2 = 0 
\lab{constrsl2}
\ee

Therefore the equations of motion and constraints of the submodel defined by
\rf{eqmovsl2} and  \rf{constrsl2} can be written as 
\be 
\pa^2 \vp - \( \pa_{\mu} \vp\)^2  =0 \; ; \qquad 
\pa^2 \zeta =0 \; ; \qquad  
\(\pa_{\mu} \vp \)^2 - e^{2 \vp} \( \pa_{\mu} \zeta \)^2 = 0 \; ; \qquad
\pa_{\mu} \zeta \pa^{\mu} \vp = 0 
\lab{subeqmov1}
\ee
Introducing
\be
\phi \equiv e^{-\vp} \; \; ; \qquad \mbox{\rm and so $\phi \geq 0$}
\ee
it becomes 
\be
\pa^2 \phi  =0 \; ; \qquad \pa^2 \zeta =0 \; ; \qquad 
\pa_{\mu} \zeta \pa^{\mu} \phi = 0  \; ; \qquad 
\( \pa_{\mu} \phi \)^2 =  \( \pa_{\mu} \zeta\)^2
\lab{subeqmov2}
\ee

Now, using the results of section \ref{sub:submodels}, we can
find an infinite number of conserved currents for the submodel
defined by equations \rf{eqmovsl2}. These currents will have
the form of \rf{curltensor2}; since for this model
\be
b = na = e^{\zeta E_{+}} \, e^{-\h \vp H} = \left( \begin{array}{lr}
e^{- \h \vp} &  \zeta e^{\h \vp} \\  0 & e^{\h \vp} \end{array} \right) \, ,
\ee
the $V_\alpha(b)$'s take the form 
\be
b \, P^{\psi} \( T_3 \) \, b^{-1} =
P^{\psi} \( \( \begin{array}{cc} 
V^{0} & V^{+} \\ V^{-} & -V^{0}
\end{array} \) \) 
\ee
with
\br
V^{+} &=& \( 1+ \zeta ^2 e^{2 \vp} \) e^{- \vp}= 
\( 1+ \frac{\zeta^2}{\phi^2}\) \phi  \nonu \\   
V^{0} &=& - \zeta e^{\vp} = -\frac{\zeta}{\phi} \nonu \\ 
V^{-} &=& - e^{\vp} = - \frac{1}{\phi} \nonu
\er

So, for the case $n=2$ in \rf{curltensor2}, one gets
\br
J_{\mu}^{2 ,(+,+)} \equiv J_{\mu}^{+} \, V^{+}  &=&  
e^{- \vp} \( 1+ \zeta^2 \, e^{2 \vp} \)
\left[ 2 \, \zeta \, \pa_{\mu} \vp + \( 1 - \zeta ^2 \, e^{2 \vp} \) \pa_{\mu}
\zeta \right] \; ; \nonu \\ 
J_{\mu}^{2, (0,0)} \equiv J_{\mu}^{0} \, V^{0}   &=& 
\zeta \, e^{\vp} \, \pa_{\mu} \vp - \zeta^2 \, e^{3
\vp} \, \pa_{\mu} \zeta \; ;\nonu \\ 
J_{\mu}^{2, (-,-)} \equiv J_{\mu}^{-} \, V^{-}  &=&  
- e^{3 \vp} \, \pa_{\mu} \zeta \; ; \nonu \\  
J_{\mu}^{2, (0,-)} \equiv  J_{\mu}^{0} \, V^{-} +  J_{\mu}^{-} \, V^{0} &=&  
e^{\vp} \pa_{\mu} \vp - 2 \,\zeta \, 
e^{3 \vp}\,  \pa_{\mu} \zeta  \; ; \nonu \\ 
J_{\mu}^{2, (0,+)} \equiv J_{\mu}^{0} \, V^{+}  + J_{\mu}^{+} \, V^{0} &=&  
2 \, \zeta^3 \, e^{3 \vp} \, \pa_{\mu} \zeta -
\( 1 + 3 \, \zeta^2 \, e^{2 \vp} \) e^{- \vp} \, \pa_{\mu} \phi  \; ; \nonu \\ 
 J_{\mu}^{2 ,(+,-)}  \equiv J_{\mu}^{+} \, V^{-} + J_{\mu}^{-} \, V^{+} &=& 
2 \, \zeta^2 \, e^{3 \vp} \, \pa_{\mu} \zeta - 
2 \, \zeta \, e^{\vp} \, \pa_{\mu} \vp \; ;
\nonu 
\er

The models discussed in section \ref{subsec:normal}, and in particular the 
example of $sl(2)$ given by \rf{eqmovsl2}, have been discussed in the
literature \ct{crem1,crem2} in the context of dualities in supergravity 
theories. It would be interesting to investigate the role of such infinite set
of conserved currents in those theories.

\sect{The $CP^N$ models}
\label{sec:cpn}

The $CP^N$ model contains $N$ complex scalar fields $u_i$, $i=1,2,\ldots N$,
and on a space-time of $d+1$ dimensions it is defined by the action
\be
S \equiv \int d^{d+1}x \, 
{\( 1 + u^{\dagger}\cdot u\) \(\pa_{\mu} \ud \cdot \pa^{\mu} u \) - 
\( \ud \cdot \pa_{\mu} u \) \( \pa^{\mu} \ud \cdot u\) 
\o {\( 1 + u^{\dagger}\cdot u\)^2}}
\lab{cpnaction}
\ee
where we have denote by $u$ the $N$-dimensional column matrix with components
$u_i$, and by $\ud$ the complex conjugate of its transpose. The corresponding
equations of motion are\footnote{Actually the equation of motion following 
from \rf{cpnaction} is $\udu \pa^2 u_i + 
2 \, {\( \ud \cdot \pa_{\mu} u\)^2 u_i \o \udu} -
2 \, \( \ud \cdot \pa_{\mu} u\) \pa^{\mu} u_i - 
\( \ud \cdot \pa^{2} u\) u_i = 0$. However, such equation (as well as 
\rf{cpneqmot}) implies, by contraction with $u^*_i$, that 
$\( \ud \cdot \pa^{2} u\) = 
2\, {\( \ud \cdot \pa_{\mu} u\)^2 \o \udu}$. Those two relations leads to 
\rf{cpneqmot}.} 
\be
\udu \pa^2 u_i = 2 \( \ud \cdot \pa_{\mu} u\) \pa^{\mu} u_i 
\lab{cpneqmot}
\ee
and the corresponding complex conjugates. 

The $CP^N$ model corresponds in fact to the non-linear sigma model on
the symmetric space $SU(N+1)/SU(N)\times U(1)$, defined in the manner 
discussed in section \ref{sec:symsp}, and therefore possesses a local zero
curvature representation as discussed there. See  \ct{fujii,lastfujii} for
alternative  
formulations.  Let $\a_i$ and $\l_i$, $i=1,2, \ldots N$,
denote the simple roots and fundamental weights respectively, of $SU(N+1)$. 
They satisfy $ {2\l_i \cdot \a_j \o \a_j^2} = \d_{ij}$. The relevant involutive
automorphism is given by
\be
\s \( T \) \equiv \Omega \, T \, \Omega^{-1} \; ; \qquad \qquad 
\Omega \equiv e^{i \pi \Lambda} \; ;\qquad \qquad 
\Lambda \equiv {2 \,\l_N \cdot H\o \a_N^2} 
\lab{involution}
\ee
with $T$ being an element of the algebra $su(N+1)$, and 
$H_i$ being the basis of its Cartan subalgebra.  
Therefore, the subalgebra of $su(N+1)$ invariant under $\s$ is generated by the
Cartan subalgebra and the step operators $E_{\pm \a}$ corresponding to roots
which are orthogonal to $\l_N$, or in other words, which do not contain $\a_N$
in its expansion in terms of simple roots. Therefore, it corresponds to the
subalgebra $su(N)\oplus u(1)$, where the simple roots of such $su(N)$ are 
the first $N-1$ simple roots of $su(N+1)$, i.e. 
$\a_a$, $a=1,2,\ldots N-1$. The $u(1)$ factor is obviously generated by
$\Lambda$ defined in \rf{involution}. Following the notation of \rf{ssdec}
one has
\br
\cs &\equiv& \{ S_{\pm i} \equiv E_{\pm \(\a_i + \a_{i+1} + \ldots \a_N\)}
\; ; \; \; i=1,2,\dots N\} \nonu \\
\ck &\equiv& su(N)\oplus u(1) 
\lab{skdeccpn}
\er

The action and equations of motion of the $CP^N$ model can then be written in 
the form \rf{sslag} and \rf{sseqmov} respectively. The main problem is to find
the correct parametrization in terms of the fields $u_i$ of the $SU(N+1)$ 
group element $g$, in \rf{pv}, such that \rf{sseqmov} reproduces
\rf{cpneqmot}. The answer to it is 
\be
g = e^{i S}\, e^{\varphi \sbr{S}{S^{\dagger}}}\, e^{i S^{\dagger}} \; \; ; 
\qquad \qquad 
\varphi \equiv {\log \sqrt{1 + u^{\dagger}\cdot u} \o {u^{\dagger}\cdot u}}
\lab{betterg}
\ee
where we have defined 
\be
S \equiv u_i \, S_i \; \; \qquad \qquad S^{\dagger} \equiv u_i^* \, S_{-i}
\lab{sdef}
\ee
with $S_{\pm i}$ introduced in \rf{skdeccpn}, and where we have used the fact
that in any finite dimensional  representation 
we can choose the basis such that $H_i^{\dagger} = H_i$ and 
$E_{\a}^{\dagger} = E_{-\a}$. 

In the $\( N+1\)$-dimensional defining 
representation of $SU(N+1)$, $g$ is given by 
\br
g \equiv {1 \o \vat}\, \( 
\begin{array}{cc}
\D & i u\\
i \ud & 1
\end{array}\) \; \; ; \qquad \qquad 
\vat \equiv \sqrt{1+ \ud \cdot u}
\lab{goodg}
\er
where $\D$ is a $N\times N$ hermitian matrix given by 
\be
\D_{ij} \equiv \vat \, \d_{ij} - {u_i u^*_j\o {1 + \vat}} \; \; ; 
\qquad \qquad i,j=1,2, \ldots N
\lab{goodd}
\ee
It then follows that $g$ is indeed an unitary matrix.

Notice that  $u$ is an eigenvector of $\D$  with unit eigenvalue
\be
\D \cdot  u = u 
\lab{eigend}
\ee

In the defining representation, $\Lambda$ and $\O$ leading to the automorphism
$\s$ of \rf{involution} are given by
\br
\Lambda = {1 \o {N+1}}\, \( 
\begin{array}{cc}
\one_{N\times N} & 0\\
0  & -N
\end{array}\)\qquad \qquad 
\O = e^{i\pi /(N+1)}\, \( 
\begin{array}{cc}
\one_{N\times N} & 0\\
0  & -1
\end{array}\)
\lab{lambdaomega}
\er
and therefore\footnote{Notice that, from \rf{involution}, \rf{betterg} and 
\rf{sigmag}, one has (in the defining representation at least) that 
$\Omega g \Omega^{-1}= e^{-i S}\, e^{\varphi \sbr{S}{S^{\dagger}}}\, 
e^{-i S^{\dagger}}= g^{-1}$, and so one can also write 
$g=  e^{i S^{\dagger}} \, e^{-\varphi \sbr{S}{S^{\dagger}}}\, e^{i S}$.}
\be
\s \( g \) = \O\, g \, \O^{-1} = g^{-1} 
\; \; \qquad \qquad \mbox{\rm and} \; \; \qquad \qquad 
y\( g\) = g^2 
\lab{sigmag}
\ee
where $y\( g\)$ is defined in \rf{pv}.

One can check that 
\br
g^{-1} \pa_{\mu} g = {1\o \vat^2} \, \( 
\begin{array}{cc}
\kappa_{\mu} & i \D \cdot \pa_{\mu} u\\
i\( \pa_{\mu} \ud \) \cdot \D & v_{\mu} 
\end{array}\) 
\lab{goodmc}
\er
where
\br
\kappa^{\mu}_{ij} &\equiv& 
{\vat \o {1+\vat}}\, \( u_i \pa^{\mu} u^*_j - \( \pa^{\mu} u_i \) u^*_j\) 
+ \h \( \ud \cdot \pa^{\mu} u - \(\pa^{\mu}\ud \) \cdot u \) 
{u_i u^*_j \o {\(1+\vat\)^2}} \nonu \\
 v_{\mu} &\equiv& \h \( \ud \cdot \pa_{\mu} u - \(\pa_{\mu}\ud \) \cdot u \)
\lab{kmu}
\er

One can write \rf{goodmc} as a linear combination of a basis  of 
$su(N+1)$ using the odd generators $S_{\pm i}$ introduced in \rf{skdeccpn}. 
In order to simplify the notation we introduce the covariant
derivative
\be
\nabla_{\mu} u_i \equiv \D_{ij} \pa_{\mu} u_j = \( \vat \, \pa_{\mu} - 
{\ud \cdot \pa_{\mu} u \o {1+\vat}} \) u_i 
\lab{covderu}
\ee

The potentials \rf{sspot2} can then be written as 
\br 
A_{\mu} &=& g^{-1}\pa_{\mu} g  \nonu\\
&=& {1 \o \vat^2}\, \( i \nabla_{\mu} S + i \(\nabla_{\mu} S\)^{\dagger} + 
{\sbr{S}{\(\pa_{\mu} S\)^{\dagger}} - 
\sbr{\pa_{\mu} S}{ S^{\dagger}} \o{1+\vat}}    
-v_{\mu} {\sbr{S}{S^{\dagger}}\o{\(1+\vat\)^2}}  \) \nonu\\
{\tilde B}_{\mu} &=& P^{\psi}\( \( 1-\s \) \( g^{-1}\pa_{\mu} g \)\) 
\lab{goodpotcpn}\\
&=& {2i \o \vat^2}\, P^{\psi}\(  \nabla_{\mu} S + 
 \(\nabla_{\mu} S\)^{\dagger} \) = 
 {2i \o \vat^2}\, \(  \nabla_{\mu} u_i \,  P^{\psi}\( S_{i}\) + 
 \(\nabla_{\mu} u_i\)^{\dagger}\, P^{\psi}\( S_{-i}\) \) \nonu
\er
Notice that in the even part  under $\s$ of $A_{\mu}$, i.e. the terms involving
commutators of $S$'s, the ordinary derivative can be replaced by the covariant
derivatives \rf{covderu} due to the antisymmetry of the terms. 

By imposing the local zero curvature condition \rf{localzc} on these
potentials one obtains the $CP^N$ equations of motion \rf{cpneqmot}. Indeed,
the flatness condition $F_{\mu\nu}=0$ is trivially satisfied since $A_{\mu}$
is of the pure gauge form. The condition $D^{\mu}{\tilde B}_{\mu}=0$ leads to
$2N$ equations which are equivalent to \rf{cpneqmot}.

According to \rf{currents} (or \rf{cosetcur}) the conserved currents of the
$CP^N$ model are given by\footnote{Where we have used the fact that in the
defining representation of $SU(N+1)$, one has 
$\( S_i\)_{rs}=\d_{ir}\d_{s,N+1}$ and 
$\( S_{-i}\)_{rs}=\d_{r,N+1}\d_{is}$, with $r,s=1,2,\ldots N+1$, and
$i=1,2,\ldots N$.}
\br
J_{\mu} &=& g \, {\tilde B}_{\mu} \, g^{-1} 
= 2\, P^{\psi} \(\( 
\begin{array}{cc}
J_{\mu}^{ij} & i J_{\mu}^i\\
i{ J_{\mu}^i}^{\dagger} & - J_{\mu}
\end{array} \) \) 
=  P^{\psi} \( J_{\mu}^{ij}\, \sbr{S_i}{S_{-j}} + i J_{\mu}^i \, S_i + 
i{J_{\mu}^i}^{\dagger}\, S_{-i}\)  \nonu\\
&=& 
{1\o {1+ \ud \cdot u}}\,  P^{\psi} \(  
\sbr{\pa_{\mu}\( S + S^{\dagger}\)}{ S + S^{\dagger}} + 
i \pa_{\mu}\( S + S^{\dagger}\) \right. \nonu\\
&-& \left. 
{\ud \cdot \pa_{\mu} u - \pa_{\mu} \ud \cdot u\o{1+ \ud \cdot u}} \, 
\( i \(  S - S^{\dagger}\) + \sbr{S}{S^{\dagger}}\) \) 
\lab{currentcpn}
\er
with $i,j=1,2, \ldots N$, and 
\br
J_{\mu}^{ij} &=& {\( 1+ \ud \cdot u\)\, \( \pa_{\mu} u_i \, \ud_{j} - u_i
\pa_{\mu} \ud_j\) - 
u_i\ud_j \( \ud \cdot \pa_{\mu} u - \pa_{\mu} \ud \cdot u \)
\o {\( 1+ \ud \cdot u\)^2}}\nonu\\
J_{\mu}^i &=& {\( 1+ \ud \cdot u\)\,  \pa_{\mu} u_i - 
u_i \( \ud \cdot \pa_{\mu} u - \pa_{\mu} \ud \cdot u \)
\o {\( 1+ \ud \cdot u\)^2}}\nonu\\
J_{\mu} &=& {\ud \cdot \pa_{\mu} u - \pa_{\mu} \ud \cdot u \o 
{\( 1+ \ud \cdot u\)^2}}
\lab{currentcpn2}
\er
In \rf{currentcpn} $J_{\mu}^{ij}$, $J_{\mu}^i$, ${ J_{\mu}^i}^{\dagger}$ and 
$J_{\mu}$ stand for matrices $(N\times N)$, $(N\times 1)$, $(1\times N)$ and
$(1\times 1)$ respectively. Notice that the number of conserved currents is
indeed equal to the dimension of $SU(N+1)$, i.e. $(N^2 +2N)$, since
$\sum_{i=1}^N J_{\mu}^{ii} =  J_{\mu}$.  

\newpage

\subsection{Integrable submodels of $CP^N$}
\label{sub:subcpn}

We now follow the strategy of section \ref{sub:submodels} to construct
submodels of $CP^N$ which presents an infinite number of conserved currents. 

Since $su(N+1)$ has no roots containing twice $\pm \a_N$ in their expansions in
terms of simple roots, it follows that $\cs$ defined in \rf{skdeccpn} splits 
into two abelian subspaces generated by $S_{i}$ and $S_{-i}$, i.e.
\be
\cs = \cs_{+} + \cs_{-}  \; ; \qquad \qquad 
\cs_{\pm} \equiv \{ S_{\pm i} \; ; \; i=1,2,\ldots N\}
\ee
and
\be
\sbr{S_i}{S_j} = \sbr{S_{-i}}{S_{-j}} = 0 \; ; \qquad \qquad 
\mbox{\rm any $i,j$}
\ee

It follows that $\cs_{+}$ and  $\cs_{-}$ transform under the representations
$N(1)$ and ${\bar N}(-1)$ respectively, of the subalgebra 
$\ck = su(N)\oplus u(1)$, i.e.
\br
\sbr{\ck}{P^{\psi}\( S_{i}\)} &=& P^{\psi}\( S_{j}\)\, R^{N(1)}_{ji}\( \ck\) 
\nonu\\
\sbr{\ck}{P^{\psi}\( S_{-i}\)} &=& P^{\psi}\( S_{-j}\)\, 
R^{{\bar N}(-1)}_{ji}\( \ck\) 
\lab{nnbar}
\er

Therefore, according to the discussion of section \ref{sub:submodels} we have
to look for representations $P^{\l}$  
of $su(N+1)$ such that its branching in
terms of  $su(N)\oplus u(1)$ possesses the representations $N(1)$ and 
${\bar N}(-1)$ at least once, i.e.
\be
P^{\l} = N(1) + {\bar N}(-1) + {\rm anything}
\lab{n1n-1}
\ee
If that happens let us denote by $P^{\l}_{i}$ and $P^{\l}_{-i}$, $i=1,2,
\ldots N$, the basis of the subspaces corresponding to $N(1)$ and 
${\bar N}(-1)$ respectively, that transform exactly like 
$P^{\psi} \( S_i\)$ and $P^{\psi} \( S_{-i}\)$, i.e. 
\br
\sbr{\ck}{P^{\l}_i} &=& P^{\l}_j\, R^{N(1)}_{ji}\( \ck\) 
\nonu\\
\sbr{\ck}{P^{\l}_{-i}} &=& P^{\l}_{-j}\, 
R^{{\bar N}(-1)}_{ji}\( \ck\) 
\lab{nnbar2}
\er
As we have commented below \rf{relrep2}, we can rescale the basis 
$P^{\l}_i$ and $P^{\l}_{-i}$ of $N(1)$ and 
${\bar N}(-1)$ respectively, independently without changing the relation 
between \rf{nnbar} and \rf{nnbar2}. Then following  \rf{goodpotcpn}, we
introduce the potential 
\be 
{\tilde B}_{\mu}^{\l} = 
 {2i \o \vat^2}\, \(  \nabla_{\mu} u_i \,  P^{\l}_i  + 
 \b \(\nabla_{\mu} u_i\)^{\dagger}\, P^{\l}_{-i} \)
\lab{lambdab}
\ee
where $\b$ is the parameter accounting for the freedom of rescaling the 
basis of the irreducible components.  

Again, according to the arguments of section \ref{sub:submodels} the zero
curvature condition
\be
D^{\mu} \, {\tilde B}_{\mu}^{\l} = 0 
\ee
where the covariante derivative is w.r.t. the same potential $A_{\mu}$ as in 
\rf{goodpotcpn}, leads to the equations of motion of the $CP^N$ model 
\rf{cpneqmot} plus the constraints (see \rf{sub2})
\br
\pa_{\mu} u_i \, \pa^{\mu} u_j \, 
\( \sbr{S_{i}}{P^{\l}_j}  + \sbr{S_{j}}{P^{\l}_i}\) &=& 0
\lab{constcpn1}\\  
\pa_{\mu} u_i^* \, \pa^{\mu} u_j^* \, 
\( \sbr{S_{-i}}{ P^{\l}_{-j}} + \sbr{S_{-j}}{ P^{\l}_{-i}}\) &=& 0  
\lab{constcpn2}\\
\pa_{\mu} u_i \,\pa^{\mu} u_j^* \,\( 
\b \sbr{S_{i}}{ P^{\l}_{-j}} + \sbr{S_{-j}}{ P^{\l}_{i}} \) &=& 0
\lab{constcpn3}
\er
In the above calculation we have used that $\nabla_{\mu} u_i = 
\D_{ij} \pa_{\mu} u_j$, together with the fact that $\D_{ij}$ is invertible,
i.e.  
\be
\D_{ij}^{-1} \equiv {1\o \vat}\, \( \d_{ij} + {u_i u^*_j\o {1 + \vat}}\)
\lab{invdelta}
\ee
Therefore, any eq. of the type 
$\nabla_{\mu} u_i \, \nabla^{\mu} u_j \, M_{ij} = 0$ 
can be written as 
$\pa_{\mu} u_i \, \pa^{\mu} u_j \, M_{ij} = 0$, for a genereic tensor
$M_{ij}$. 

We have that the terms involving commutators in \rf{constcpn1}, 
\rf{constcpn2}, and \rf{constcpn3} transform under $\ck = su(N)\oplus u(1)$ as 
$\( N\times N\)_s (2) = {N(N+1)\o 2} (2)$, 
$\( {\bar N}\times {\bar N}\)_s (-2) = {N(N+1)\o 2} (-2)$, and 
$\( N\times {\bar N}\) (0) = \( 1 + {\rm adjoint}\) (0)$, respectively. 
Therefore, as 
we discussed in section \ref{sub:submodels}, the constraints 
\rf{constcpn1}-\rf{constcpn3} will only be 
effective if such representation appear in the branching of $P^{\l}$ 
in terms of representations of $\ck = su(N)\oplus u(1)$. 

In any case, the model defined by the equations \rf{cpneqmot} and constraints 
\rf{constcpn1}-\rf{constcpn3} possesses the conserved currents (see
\rf{currents}) 
\be
J^{\l}_{\mu} \equiv  g\, {\tilde B}_{\mu}^{\l} \, g^{-1} 
\lab{currentslambda}
\ee
with $g$ given by \rf{betterg}. If the number of representations 
$P^{\l}$ satisfying
the conditions discussed above is infinite, one obviously gets an infinite 
number of conserved currents. 

\subsubsection{The singlet states and infinite number of currents}
\label{subsub:singletcpn}

The adjoint representation of $SU(N+1)$ decomposes into representations of
$SU(N)\otimes U(1)$ as
\be
Adj \( SU(N+1)\) = N(1) + {\bar N}(-1) + Adj \( SU(N)\)(0) + 1(0)
\ee
Therefore it possesses a singlet state satisfying \rf{singletcond}. That
singlet corresponds to the $U(1)$ generator $\Lambda$ defined in
\rf{involution} and \rf{lambdaomega}. 

Consequently, we can apply the ideas of section \ref{sub:singlet} to construct
an infinite number of conserved currents for submodels of $CP^N$. Denoting the
generators of $\ck = su(N)\oplus u(1)$, by $K_r$, $r=1,2,\ldots N^2$, one can
write the potential $A_{\mu}$ in \rf{goodpotcpn} as 
\be
A_{\mu} = i\, \( A_{\mu}^r K_r +  b_{\mu} +  b^{\dagger}_{\mu}\) \; ; \qquad 
b_{\mu} \equiv { \nabla_{\mu} S \o {1 + \ud \cdot u}}
\ee
Then, following \rf{cosetpottensor} we define 
\br
A_{\mu}^{(n)} &\equiv&  i\, \sum_{l=0}^{n-1} 
\(\otimes 1\)^l \otimes \( A_{\mu}^r K_r +  b_{\mu} +  b^{\dagger}_{\mu} \) 
\(\otimes 1\)^{n-l-1} 
\lab{cpnpottensor}\\
{\tilde B}_{\mu}^{\psi (n)} &\equiv&  2i \; 
\sum_{l=0}^{n-1}
\;  \(\otimes  P^{\psi}\(\Lambda\)\)^l \otimes 
 P^{\psi} \( c_{n,l}\, b_{\mu} + {\bar c}_{n,l}\, b^{\dagger}_{\mu} \)  
\(\otimes  P^{\psi}\(\Lambda\)\)^{n-l-1}
\nonu
\er
Since the representation $R^{\cs}= N(1) + {\bar N}(-1)$, is reducible we can
rescale each irreducible component independently (see comments below
\rf{relrep2} and \rf{cosetpottensor}). The constants $c_{n,l}$ and 
${\bar c}_{n,l}$ account for such freedom. 

We now impose that these potentials should satisfy the zero curvature
conditions \rf{zclambda}. Obviously $A_{\mu}^{(n)}$ satisfy $F_{\mu\nu}=0$. As
we have argued in section \ref{sub:singlet}, the
components of the condition $D^{\mu} {\tilde B}_{\mu}^{\psi (n)}=0$, involving
$\pa^{\mu} {\tilde B}_{\mu}^{\psi (n)}$ and the commutator of 
${\tilde B}_{\mu}^{\psi (n)}$ with the $\ck$-part of $A_{\mu}^{(n)}$  lead to
the equations of motion of the $CP^N$ 
model \rf{cpneqmot}.  The commutator of 
${\tilde B}_{\mu}^{\psi (n)}$ with the $\cs$-part of $A_{\mu}^{(n)}$ 
leads to the constraints defining the submodel. We analyze those constraints by
collecting the linearly independent terms in the tensor product. The terms
involving $b_{\mu}$'s in the $l$ and $m$ positions 
of the tensor product are ($l<m$) 
\br
&& \( c_{n,l} + c_{n,m}\) \(\otimes  P^{\psi}\(\Lambda\)\)^{l-1} 
\otimes  P^{\psi} \(  b_{\mu}  \) 
\(\otimes  P^{\psi}\(\Lambda\)\)^{m-l} 
 P^{\psi} \(  b^{\mu}  \)  
\(\otimes  P^{\psi}\(\Lambda\)\)^{n-m}  
\lab{constrainttensor1}\\
&-& \( {\bar c}_{n,l} + {\bar c}_{n,l}\)\(\otimes  P^{\psi}\(\Lambda\)\)^{l-1} 
\otimes  P^{\psi} \(  b^{\dagger}_{\mu} \) 
\(\otimes  P^{\psi}\(\Lambda\)\)^{m-l} 
 P^{\psi} \( {b^{\dagger}}^{\mu}  \)  
\(\otimes  P^{\psi}\(\Lambda\)\)^{n-m}  \nonu\\
&-& \( c_{n,l} - {\bar c}_{n,m}\)\(\otimes  P^{\psi}\(\Lambda\)\)^{l-1} 
\otimes  P^{\psi} \(  b^{\mu} \) 
\(\otimes  P^{\psi}\(\Lambda\)\)^{m-l} 
 P^{\psi} \( b^{\dagger}_{\mu}  \)  
\(\otimes  P^{\psi}\(\Lambda\)\)^{n-m}  \nonu\\
&+& \( {\bar c}_{n,l} - c_{n,m}\)\(\otimes  P^{\psi}\(\Lambda\)\)^{l-1} 
\otimes  P^{\psi} \(  b^{\dagger}_{\mu} \) 
\(\otimes  P^{\psi}\(\Lambda\)\)^{m-l} 
 P^{\psi} \( b^{\mu}  \)  
\(\otimes  P^{\psi}\(\Lambda\)\)^{n-m} = 0 \nonu
\er
where we have used the fact that $\sbr{\Lambda}{b_{\mu} +  b^{\dagger}_{\mu}}
=  b_{\mu} -  b^{\dagger}_{\mu}$. 

The terms involving commutators of $b_{\mu}$'s are 
\br
\( c_{n,l} - {\bar c}_{n,l}\)\(\otimes  P^{\psi}\(\Lambda\)\)^{l} 
\otimes  P^{\psi} \(  \sbr{b^{\mu}}{b^{\dagger}_{\mu}} \)   
\(\otimes  P^{\psi}\(\Lambda\)\)^{n-l-1} = 0 
\lab{constrainttensor2}
\er

Therefore, if we choose 
\be
c_{n,l}={\bar c}_{n,l}=1 \; ; \qquad \qquad \mbox{\rm for any $n$ and $l$} 
\lab{nicechoice}
\ee 
we get the contraints $\nabla_{\mu} u_i \, \nabla^{\mu} u_j = 0$ and 
$\(\nabla_{\mu} u_i\)^{\dagger} \, \(\nabla^{\mu} u_j\)^{\dagger} = 0$, for any
$i$ and $j$. However, from \rf{covderu} we have that $\nabla_{\mu} u_i = 
\D_{ij} \pa_{\mu} u_j$, and since $\D_{ij}$ is invertible (see \rf{invdelta}), 
it follows that the constraints become just 
$\pa_{\mu} u_i \, \pa^{\mu} u_j = 0$ and their complex conjugates. Using such
contraints on the equation of motion of $CP^N$ \rf{cpneqmot} one gets that 
$\pa^2 u_i =0$. 
Consequently, the submodel we obtain is defined by the equations
\be
\pa^2 u_i =0 \; ; \qquad \qquad \pa_{\mu} u_i \, \pa^{\mu} u_j = 0 
\lab{nicesub}
\ee
and the corresponding complex conjugate equations. 

According to the discussions of section \ref{sub:singlet} such submodel
possesses an infinite number of currents given by \rf{curltensor2}. The
quantities $V_{\a}$, defined in \rf{vdefsing}, are given by 
\br
g\, P^{\psi}\( \Lambda \) g^{-1} &=& P^{\psi}\( \(
\begin{array}{cc}
V^{ij} & -i V^{i}\\
i {V^{i}}^{\dagger}& -V
\end{array}\) \) 
=  P^{\psi}\( V^{ij}\, \sbr{S_i}{S_{-j}} - i V^{i}\, S_i 
+ i {V^{i}}^{\dagger}\, S_{-i} \)
\nonu\\
&=&  P^{\psi}\( \Lambda - {1\o {1 + \ud \cdot u}}\, \( 
\sbr{S}{S^{\dagger}} + i \( S - S^{\dagger}\)\) \) 
\lab{vdefcpn1}
\er
with  $i,j =1, 2, \ldots N$, and  
\br
V^{ij} &\equiv& {\d_{ij}\o N+1}- {u_i \ud_j\o {1 + \ud \cdot u}} \nonu\\
V^i &\equiv& {u_i \o {1 + \ud \cdot u}} \nonu\\
V  &\equiv& -{1\o N+1} + {1\o {1 + \ud \cdot u}} 
\lab{vdefcpn2}
\er
In \rf{vdefcpn1} we are using the same notation as in \rf{currentcpn}. The
number of independent quantities $V$'s is the dimension of $SU(N+1)$, since
$\sum_{i=1}^N V_{ii} = V$. 

One can easily check that the currents \rf{currentcpn2} can be written in
terms of \rf{vdefcpn2} as
\br
J^{ij}_{\mu} &=& -\( {\d V^{ij} \o \d u_m } \pa_{\mu}
u_m - {\d V^{ij} \o \d \ud_m } \pa_{\mu} \ud_m \) \nonu \\
J^{i}_{\mu} &=& {\d V^{i} \o \d u_m } \pa_{\mu}
u_m - {\d V^{i} \o \d \ud_m } \pa_{\mu} \ud_m \nonu \\
{J^{i}_{\mu}}^{\dagger} &=& -\( {\d {V^{i}}^{\dagger} \o \d u_m } \pa_{\mu}
u_m - {\d {V^{i}}^{\dagger} \o \d \ud_m } \pa_{\mu} \ud_m \)
\er

Since we have chosen all the $c_{n,l}$'s to be unity (see \rf{nicechoice}), it
follows that all the conserved currents \rf{curltensor2} are of the form
\be
J_{\mu}^{\a_1\a_2\ldots \a_n} \equiv  
{\d {\cal F}^{\a_1\a_2\ldots \a_n} \o \d u_m } \pa_{\mu} u_m 
- {\d {\cal F}^{\a_1\a_2\ldots \a_n} \o \d \ud_m } \pa_{\mu} \ud_m  
\lab{currentsncpn}
\ee
where  
\be
{\cal F}^{\a_1\a_2\ldots \a_n} \equiv - \prod_{l=1}^n \, V^{\a_l}
\ee
with $V^{\a_l}$'s being any of the quantities in \rf{vdefcpn2}, i.e 
$V^{ij}$, $V^i$ and ${V^i}^{\dagger}$.

In fact, any quantity of the form 
\be
J_{\mu} = {\cal M}_i \pa_{\mu} u_i + {\cal N}_i \pa_{\mu} \ud_i 
\ee
with ${\cal M}_i$ and ${\cal N}_i$ being functionals of $u_j$ and $\ud_j$, 
is a conserved current of the submodel \rf{nicesub} provided \ct{lastfujii} 
\be 
{\d {\cal M}_i \o \d \ud_j} + {\d {\cal N}_j \o \d u_i}=0
\ee
The conserved quantitites we have obtained above, using the ideas of section
\ref{sub:singlet}, are particular examples of the cases where 
${\cal M}_i = {\d {\cal F} \o \d u_i}$ and  
${\cal N}_i = - {\d {\cal F} \o \d \ud_i}$. 

It is worth mentioning that the amount of conservation laws the submodel
\rf{nicesub} possesses is due (at least partially) to the huge symmetry group
it presents. Indeed, the submodel \rf{nicesub} is invariant under the
transformations 
\br
u_i \ra u_i^{\pr} &\equiv& \omega_i^{(1)} \( u\) + \omega_i^{(2)} \( u^*\)
\nonu\\
u_i^* \ra u_i^{*\,\pr} &\equiv& 
\omega_{-i}^{(1)} \( u\) + \omega_{-i}^{(2)} \( u^*\) 
\lab{hugesymm}
\er
provided
\br
\omega_i^{(1)}\( u\) \omega_j^{(2)}\( u^*\) + 
\omega_i^{(2)}\( u^*\) \omega_j^{(1)}\( u\) &=& h_{ij}^{(1)}\( u\) +
h_{ij}^{(2)}\( u^*\) \nonu\\
\omega_{-i}^{(1)}\( u\) \omega_{-j}^{(2)}\( u^*\) + 
\omega_{-i}^{(2)}\( u^*\) \omega_{-j}^{(1)}\( u\) &=& h_{-i,-j}^{(1)}\( u\) +
h_{-i,-j}^{(2)}\( u^*\)  
\lab{hugecond}
\er
for any $i,j,k,l=1,2,\ldots N$, and 
where $\( \omega_{\pm i}^{(1)}, h_{\pm i,\pm j}^{(1)}\)$ and 
$\( \omega_{\pm i}^{(2)},  h_{\pm i,\pm j}^{(2)}\)$
are functions of $u_i$'s and $u_i^*$'s only,
respectively. Particular 
solutions for \rf{hugecond} are obtained by taking 
$\omega_{i}^{(1)} = {\rm const.}$, and $\omega_{j}^{(2)}$ arbitrary, or
vice-versa.  The same being true for the negative $\omega$'s. 
Therefore, starting with a suitable small set of currents, a large amount 
of other currents can be construct using  the transformations
\rf{hugesymm}. 

\newpage 

In the case $n=2$, the currents \rf{currentsncpn} are given by 
\br
J_{\mu}^{ij}V^{kl} + V^{ij}J_{\mu}^{kl}
&=& {1 \o {\( 1+ \ud \cdot u\)^2}}
\left\{ {2 (\ud \cdot \pa_{\mu} u - \pa_{\mu} \ud \cdot u) \o 1+ \ud \cdot u}
u_i \ud_j u_k \ud_l \right. \nonu \\ &+& \left.
{( 1 + \ud \cdot u ) \o (N + 1)} \left[
\d _{kl}( \pa_{\mu} u_i \ud_{j}
-  u_i \pa_{\mu} \ud_{j}) + \d _{ij}(\pa_{\mu} u_k \ud_{l} - 
u_k \pa_{\mu} \ud_{l}) \right] \right. \nonu \\ 
&-& \left. u_i \ud_{j} ( \pa_{\mu} u_k \ud_{l} - u_k \pa_{\mu} \ud_{l} )
- u_k \ud_{l} ( \pa_{\mu} u_i \ud_{j} - u_i \pa_{\mu} \ud_{j} ) \right. \nonu
\\ &-& \left.{ (\ud \cdot \pa_{\mu} u - \pa_{\mu} \ud \cdot u) \o N + 1}
\( \d _{kl} u_i \ud_j + \d_{ij} u_k \ud_l \) 
\right\} \nonu\\
J_{\mu}^{ij}V^{k} - V^{ij}J_{\mu}^{k} 
&=& {1 \o {\( 1+ \ud \cdot u\)^2}} \left[
u_k \( \pa_{\mu} u_i \ud_{j} - u_i \pa_{\mu} \ud_{j} \) + u_i \ud_j \pa_{\mu}
u_k  \right. \nonu \\ &-& \left. {2 \( \ud \cdot \pa_{\mu} u -
\pa_{\mu} \ud \cdot u \) \o 1 + \ud \cdot u} u_i \, \ud_j \, u_k 
\right. \nonu \\ &-& \left. { 1 + \ud \cdot u \o N + 1} \d_{ij} 
\pa_{\mu} u_k + { (\ud \cdot \pa_{\mu} u - \pa_{\mu} \ud \cdot u) \o N + 1}
u_k \d_{ij} \right] \nonu\\
J_{\mu}^{i}V^{j} + V^{i}J_{\mu}^{j} 
&=&  {u_i \pa_{\mu} u_j + u_j \pa_{\mu} u_i \o {\( 1+ \ud \cdot u\)^2}}
- {2 u_i u_j \o \( 1 + \ud \cdot u\)^3} 
\( \ud \cdot \pa_{\mu} u - \pa_{\mu} \ud \cdot u \)
\nonu\\
J_{\mu}^{ij}{V^{k}}^{\dagger} + V^{ij}{J_{\mu}^{k}}^{\dagger} 
&=& {1 \o {\( 1+ \ud \cdot u\)^2}} \left[ \ud_k (\pa_{\mu} u_i \ud_{j} - u_i
\pa_{\mu} \ud_{j} ) \right. \nonu \\ &-& \left. 2 { (\ud \cdot \pa_{\mu} u -
\pa_{\mu} \ud \cdot u) \o  1 + \ud \cdot u} u_i \ud_j \ud_k - u_i \ud_j
\pa_{\mu} \ud_k \right. \nonu \\ &+& \left. { 1 + \ud \cdot u \o N + 1}
\d_{ij} \pa_{\mu} \ud_k + { (\ud \cdot \pa_{\mu} u - \pa_{\mu} \ud \cdot u) \o
 N + 1} \d_{ij} \ud_k \right]  
\nonu \\
{J_{\mu}^{i}}^{\dagger}V^{j} - {V^{i}}^{\dagger}J_{\mu}^{j}
 &=& { 1 \o { \( 1 + \ud \cdot u  \)^3 }}
\left\{ \( 1 + \ud \cdot u \) 
\( \pa_{\mu} \ud_{i} u_j - \ud_{i} \pa_{\mu} u_j \) \right. \nonu \\ 
&+& \left. 2 \, \ud_i u_j \( \ud \cdot \pa_{\mu} u - \pa_{\mu} \ud \cdot u  \)
\right\} \lab{currentsn=2} \\  
{J_{\mu}^{i}}^{\dagger}{V^{j}}^{\dagger} 
+ {V^{i}}^{\dagger}{J_{\mu}^{j}}^{\dagger} &=&
{\pa_{\mu} \ud_{i} \ud_{j} + \ud_{i} \pa_{\mu} \ud_{j}\o 
{\( 1+ \ud \cdot u\)^2}} 
- {2 \ud_i \ud_j \o \( 1 + \ud \cdot u\)^3} 
\( \pa_{\mu} \ud \cdot u - \ud \cdot \pa_{\mu} u \)  
\nonu 
\er

Notice that with the choice \rf{nicechoice}, the operator 
${\tilde B}_{\mu}^{\psi (n)}$ given in \rf{cpnpottensor}, belongs to the
symmetric part of the tensor product. Therefore, the above currents are
associated to the irreducible representations of $SU(N+1)$ in the symmetric
part of the tensor
product of the adjoint representation with itself. For instance, in the case
of $N=1$ one gets that the adjoint (triplet) of $SU(2)$ satisfies 
${\bf 3}\otimes {\bf 3} = {\bf 5}_s + {\bf 3}_a + {\bf 1}_s$. Indeed, 
for $N=1$ one can easily check that \rf{currentsn=2} gives $6$ currents and
that $5$ of them  coincide with the spin $j=2$ (${\bf 5}$) 
currents calculated in  \ct{afg} for the example of $CP^1$. The sixth one
coincides with one of the spin $j=1$ currents of \ct{afg}.

\subsubsection{Further submodels}

When analyzing the constraints \rf{constrainttensor1} and
\rf{constrainttensor2} we looked for a solution valid for any $n$ in order to
have an infinite number of currents. However, for the case $n=2$ there is an
additional solution, besides \rf{nicechoice}, which corresponds to
\be
c_{2,0} + c_{2,1}=0 \; \; ; \qquad {\bar c}_{2,0} + {\bar c}_{2,1}=0
\ee
Such choice leads to the submodel of $CP^N$ defined by 
\be
\udu \pa^2 u_i = 2 \( \ud \cdot \pa_{\mu} u\) \pa^{\mu} u_i \; \; ; 
\qquad \qquad \pa_{\mu}u_i \, \pa^{\mu}u^*_j = 0
\lab{cpneqmotsub}
\ee
with $i,j=1,2, \ldots N$. Therefore, using the same procedures of section
\ref{subsub:singletcpn} one obtains conserved currents of the type
\rf{currentsn=2}. In such case, the currents will depend upon one parameter
which is the ratio ${\bar c}_{2,0}/c_{2,0}$. 

Additional conserved  currents for the submodel \rf{cpneqmotsub} can be
constructed using the ideas of section \ref{sub:subcpn}. As an example
consider the case of $CP^2$. The representations ${\bf 10}$ and  
${\bar{\bf 10}}$ of $SU(3)$ break into irreducibles of $SU(2)\otimes U(1)$ as
\br
{\bf 10} &=& {\bf 4} (1) + {\bf 3} (0) + {\bf 2} (-1) + {\bf 1} (-2)\nonu\\
{\bar{\bf 10}} &=& {\bf 4} (-1) + {\bf 3} (0) + {\bf 2} (1) + {\bf 1} (2)
\er
Therefore, ${\bf 10}+{\bar{\bf 10}}$ contains the representation 
$R^{\cs}={\bf 2} (1) + {\bf 2} (-1)$ discussed in \rf{n1n-1}, and therefore
one can defined an operator ${\tilde B}_{\mu}$ like in \rf{lambdab} using such
representations. One can check
that the constraints \rf{constcpn1}-\rf{constcpn3} can be solved by imposing 
$\pa_{\mu}u_i \, \pa^{\mu}u^*_j = 0$, $i,j=1,2$. Then through
\rf{currentslambda} one obtains $20$ conserved currents for the corresponding 
submodel \rf{cpneqmotsub} of $CP^2$. A more careful analysis is necessary to
work out all the conserved currents of such type of submodels.

\vspace{3 cm}

\noindent {\bf Acknowledgements} \\
We are grateful to J.A.C. Alcaraz, H. Aratyn, J.F. Gomes and A.H. Zimerman for
interesting discussions. L.A.F. is partially supported by CNPq (Brazil) and
E.E.L. is supported by FAPESP (Brazil).

\end{document}